\documentclass[10pt,runningheads]{llncs}

\usepackage[T1]{fontenc}
\usepackage{amsmath,amsfonts}
\usepackage{algorithmic}
\usepackage{algorithm}
\usepackage{array}
\usepackage{subfig}
\usepackage{textcomp}
\usepackage{stfloats}
\usepackage{url}
\usepackage{verbatim}
\usepackage{graphicx}
\usepackage{cite}

\usepackage {hyperref}
\usepackage{booktabs}
\usepackage{colortbl}
\usepackage{rotating}
\usepackage{multirow}
\usepackage{lscape}
\usepackage{tikz}

\usepackage{libertine}

\usepackage{graphicx}
\begin{document}
\newcommand{\state}[1]{\textit{#1}}
\title{Externally validating the IoTDevID device identification methodology using the CIC IoT 2022 Dataset\thanks{Kahraman Kostas supported by Republic of Turkey - Ministry of National Education}}

\titlerunning{Externally validating the IoTDevID methodology using the CIC IoT 2022 dataset}
%
%
\author{Kahraman Kostas\inst{1}\orcidID{0000-0002-4696-1857} \and
Mike Just\inst{1}\orcidID{0000-0002-9669-5067} \and
Michael A. Lones\inst{1}\orcidID{0000-0002-2745-9896}}
\authorrunning{K. Kostas et al.}
%
\institute{Department of Computer Science, Heriot-Watt University, Edinburgh EH14 4AS, UK \\
\email{\{kk97,m.just,m.lones\}@hw.ac.uk}}
\maketitle           
\begin{abstract}

In the era of rapid IoT device proliferation, recognizing, diagnosing, and securing these devices are crucial tasks. The IoTDevID method (IEEE Internet of Things ’22) proposes a machine learning approach for device identification using network packet features.
In this article we present a validation study of the IoTDevID method by testing core components, namely its feature set and its aggregation algorithm, on a new dataset. 
The new dataset (CIC-IoT-2022) offers several advantages over earlier datasets, including a larger number of devices, multiple instances of the same device, both IP and non-IP device data, normal (benign) usage data, and diverse usage profiles, such as \state{active} and \state{idle} states. Using this independent dataset, we explore the validity of IoTDevID's core components, and also examine the impacts of the new data on model performance. 
Our results indicate that data diversity is important to model performance. For example, models trained with \state{active} usage data outperformed those trained with \state{idle} usage data, and multiple usage data similarly improved performance. 
Results for IoTDevID were strong with a 92.50 F1 score for 31 IP-only device classes, similar to our results on previous datasets. In all cases, the IoTDevID aggregation algorithm improved model performance. For non-IP devices we obtained a 78.80 F1 score for 40 device classes, though with much less data, confirming that data quantity is also important to model performance.

\end{abstract}

\section{Introduction}

An internet of things (IoT) device can be defined as any kind of physical device with processing capability that can be connected to the internet or other devices~\cite{IoT2023What}. 
Today, the number of IoT devices has exceeded 10 billion and is expected to reach 27 billion by 2025\cite{IoT2023sta}. In a rapidly growing market, a variety of devices have been developed by many companies for many purposes in a short time. 
Due to their various uses and physical requirements, these devices have very different hardware and software characteristics.  
The heterogeneity of these devices,  along with inherent vulnerabilities introduced by manufacturers and the presence of unfamiliar device interfaces, renders them susceptible to potential security risks.
Research indicates that
an IoT device connected to the internet is attacked within 5 minutes and becomes the target of a specialised attack within 24 hours~\cite{modi_2019}.

To cope with these attacks, it is essential to keep the devices up-to-date, identify the vulnerabilities they carry and find solutions for them. These devices may need to be updated, restricted or isolated from other devices depending on their vulnerabilities. In any measure to be taken, the first step will be to identify the device. 
However, the heterogeneous structure of IoT devices makes the device identification process challenging. In this regard, many researchers are applying machine learning-based identification for more efficient solutions. 

While several such studies exist, they often suffer from methodological issues that affect the reliability of their results, including data leakage, feature overfitting and selective device testing. 
We previously
created IoTDevID~\cite{kostas2022iot} to address the device identification problem, while following sound methodological principles.  
IoTDevID works at the individual packet level to identify IoT devices, whether IP or non-IP (such as Z-Wave, ZigBee, or Bluetooth). In doing so, it provides a high detection rate thanks to its incorporated aggregation algorithm, 
which 
combines similarly-modelled packets and 
improves identification success over using individual packets. 
In the multi-layer feature selection process, device and session-based identifying features that cause overfitting are discarded, and the most appropriate feature set is created by using a genetic algorithm. We further performed training and testing on isolated datasets in order to eliminate data leakage issues. In this context, IoTDevID claimed to provide generalisable and robust models.  

In this study, we validate our IoTDevID solution by applying it to a new dataset, the \href{https://www.unb.ca/cic/datasets/iotdataset-2022.html}{CIC IoT Dataset 2022} (CIC-IoT-22). 
This dataset provides an opportunity to 
test the robustness and generalisability of core components of our solution, namely its feature set and aggregation algorithm. CIC-IoT-22 
contains more devices than other prominent datasets used in our original evaluation of IoTDevID: Aalto~\cite{miettinen2017iot,aalto2017dataset} and UNSW~\cite{sivanathan2018classifying}. 
It also has non-IP devices (which UNSW does not) and data collected during use (which Aalto does not). It also contains additional contextual data, such as whether a device is \state{idle} or \state{active}, as well as different data usage scenarios. This richer dataset will allow us to further test the generalisability of IoTDevID, and it also allows us to provide some insight into the usefulness of such data for creating more generalisable and robust models. In order to enhance transparency and ensure reproducibility, we have publicly shared our dataset, and scripts\footnote{Complete feature list:\href{https://github.com/kahramankostas/IoTDevID-CIC/}{github.com/kahramankostas/IoTDevID-CIC}} .



\section{Related Work on Device Identification}\label{LR}

Device identification aims to classify devices by using feature sets (fingerprints) obtained from network data as input.  These features are usually derived from individual packet headers or payloads~\cite{kostas2022iot,miettinen2017iot,CIC,aksoy2019automated,bezawada2018behavioral}, but some studies have also used flow features~\cite{hamad2019iot,sivanathan2018classifying}. Although much work has been done in the area of device identification, a number of problems are apparent, 
including
data leakage, overly-specific features, selective device testing, and 
insufficiently transparent experimental methodology.
As in many security~\cite{arp2022and} and machine learning studies~\cite{kapoor2022leakage}, reproducibility is a serious problem in device identification. The major factor causing the reproducibility problem in device identification is data leakage. This is often caused by an improper separation of testing and training data. 
For example, in  Chowdhury et al.'s study~\cite{chowdhury2023device}, during feature extraction, features that could uniquely identify sessions (e.g. port numbers, TCP sequence, and TCP acknowledgement) were used.
Since data sessions were not considered when splitting training and testing data, data leakage from the training data would very likely cause an overestimation of a model's performance on the test dataset. 


Similarly, in the IoTSentinel~\cite{miettinen2017iot} study, models are trained with the IP address count feature which is dependent on the number of device communications in the network.
However, this is primarily determined by the network to which the device is connected rather than the device itself. Consequently, this feature is not generalisable as it will change when the device or model is moved to another network.
%
Additionally, Hamad et al.\@~\cite{hamad2019iot} used 67 features consisting of network statistics derived from 20-21 consecutive individual packet features. However, these statistics are specific to the network in which they are produced.  If the same device or model is moved to another network, these network statistics will change and the model will no longer function. 
As a further example, Sivanathan et al.\@\cite{sivanathan2018classifying},  include similarly network-dependent flow-based features 
 to create their models. 

A further
problem is that  
several studies suffer from a
lack of transparency, which is important for experiment validation and repeatability.
For example, IoTSense~\cite{bezawada2018behavioral} discarded four out of 14 devices during the evaluation step. Aksoy et al.\@~\cite{aksoy2019automated} used only 23 devices of the Aalto dataset, which has 27 devices.  Sivanathan  et al.\@\cite{sivanathan2018classifying} similarly did not include the four devices in their dataset in their results. 
Partial device use, especially when done without adequate motivation, undermines the reliability of results.
In a similar way, the IoTsense~\cite{bezawada2018behavioral} dataset has not been shared and, as far as we are aware, no code from any of the above studies~\cite{miettinen2017iot,aksoy2019automated,bezawada2018behavioral,sivanathan2018classifying} has been made publicly available. In such cases, full study validation and repeatability are not possible.

\begin{figure}[ht]
	\centerline{\includegraphics[width=.60\columnwidth]{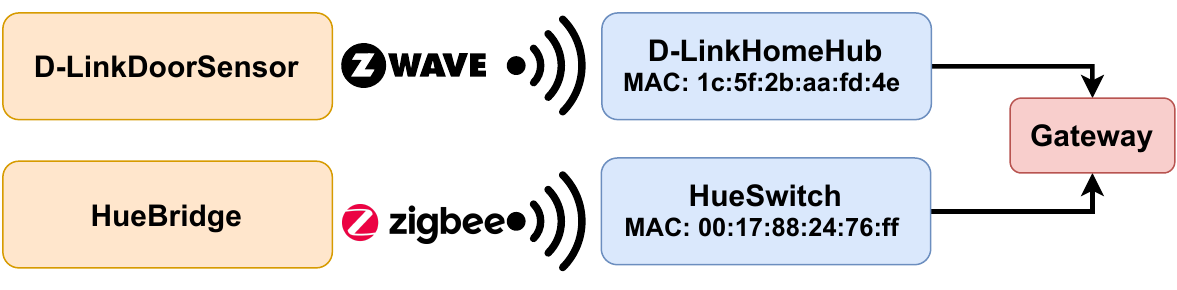}}
	\caption{
 Example
 transfer problem in the Aalto dataset. Network data is collected at the gateway.  Between D-LinkDoorSensor and D-LinkHomeHub there is only Z-Wave as a communication medium. Between D-LinkHomeHub  and the gateway, there is only Ethernet. Data from the D-LinkDoorSensor  is decapsulated at the D-LinkHomeHub and re-encapsulated for transmission to the gateway. As a result, both the  D-LinkDoorSensor and D-LinkHomeHub  share the same MAC address as the data is encapsulated on the same device (the same happens between HueBridge, HueSwitch and gateway). Therefore, studies that use MAC addresses to concatenate packets cannot overcome this problem.}
	\label{fig:transfer}
\end{figure}

Another issue is the \textit{transfer problem} (see Fig.~\ref{fig:transfer}), which impacts studies that combine individual packets or use flows. Even though many studies use individual packets, they combine these packets using features such as MAC or IP addresses. Unfortunately, they cannot solve the case where MAC/IP addresses represent more than one device. For example, they mistakenly assume that separate devices with the same IP gateway address are the same device. Among the studies using the Aalto dataset, IoTsentinel~\cite{miettinen2017iot} suffers from transfer problems because it uses MAC addresses and \cite{hamad2019iot} uses IP addresses. On the other hand, UNSW and IoTSense datasets do not have non-IP devices, so they do not have this problem, but their feature extraction method does not incorporate solutions for the transfer problem.
We offered a solution to the transfer problem as part of our aggregation algorithm in IoTDevID, which we describe below.

\subsection{IoTDevID} \label{IoTDevID} 

\begin{figure}[ht]
	\centerline{\includegraphics[width=1.0\columnwidth]{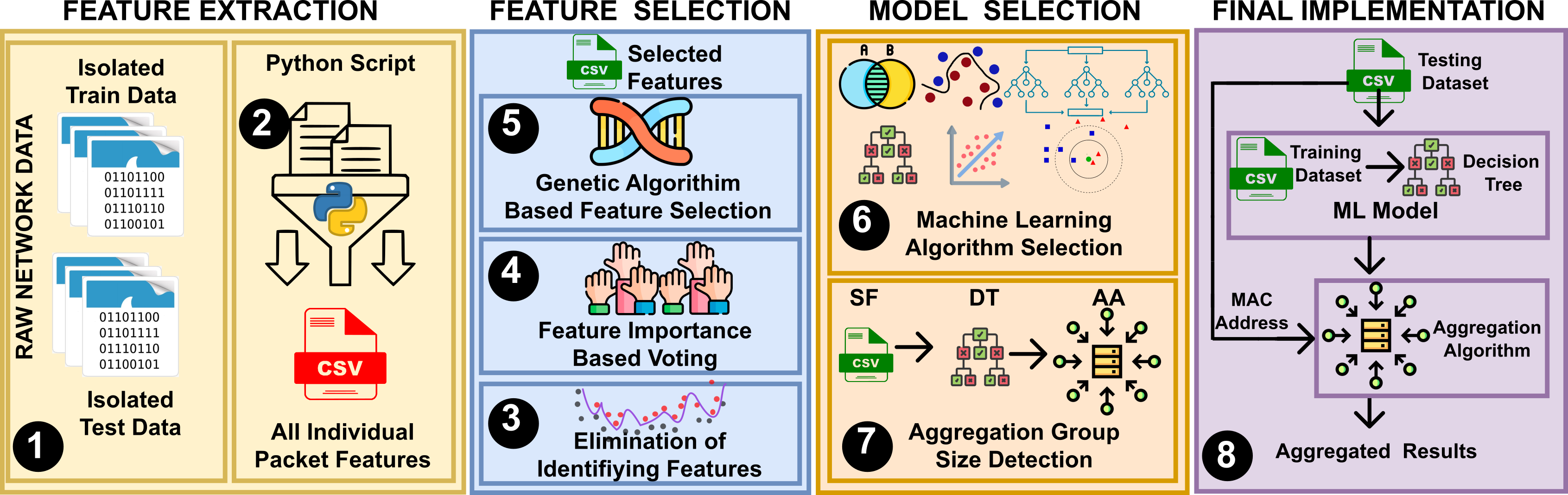}}
	\caption{Steps of the IoTDevID study.}
	\label{fig:IoT-DevID}
\end{figure}

Fig.~\ref{fig:IoT-DevID} shows the steps of the IoTDevID~\cite{kostas2022iot} study. With reference to this figure, IoTDevID can be summarised as follows:  Network data are isolated from each other by separating them into training and testing \textcircled{\small{1}}. Individual packet features are extracted from the isolated pcap files \textcircled{\small{2}}.  Identifying features that could cause overfitting were identified and discarded \textcircled{\small{3}}.
Using a voting method based on feature importance scores, unimportant features are eliminated \textcircled{\small{4}}. From the remaining features, a genetic algorithm is used to find the best feature combination \textcircled{\small{5}}. Different machine learning algorithms are tested to find the most appropriate algorithm \textcircled{\small{6}}.  The optimal size for the aggregation algorithm is determined \textcircled{\small{7}}. In the last step, the final results are obtained by using the ML model and the aggregation algorithm \textcircled{\small{8}}.

In the IoTDevID study, we aimed for a transparent, repeatable and generalisable study, avoiding the pitfalls found in previous studies, such as data leakage, use of identifying features, selective device testing, and non-transparent experiments. We used features derived from packet headers as used in many other studies~\cite{miettinen2017iot,CIC,aksoy2019automated,bezawada2018behavioral}. However, device identification with individual packets is very difficult due to the high noise.  This noise is caused by the fact that some ``empty'' packets have multiple device characteristics. An example of this is the TCP  3-way handshake. For this handshake, only empty packets with the TCP flags 
are sent. These packets are quite simple and stable/static. It is very difficult to tell from a single packet which device it came from once identifying data such as IP/MAC addresses is removed. Therefore mislabelling of the fingerprint from these packets is quite common. To combat this, some researchers\cite{miettinen2017iot,bezawada2018behavioral} have constructed more descriptive fingerprints by combining features from successive packets. 
The problem with this approach is that since the combination process uses identifying features such as MAC/IP, it does not work in networks where there are transfer problems or non-IP devices.

\begin{figure}[ht]
	\centerline{\includegraphics[width=.70\columnwidth]{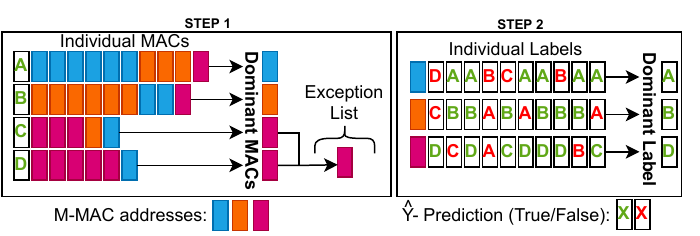}}
	\caption{The steps of the aggregation algorithm. In the first step, MAC addresses are sorted by tags to find MAC addresses that show more than one device behaviour and added to the exception list. In the second step, labels outside the exception list are grouped according to MAC addresses. The most repetitive label is assigned to the whole group and the aggregated results are created.}
	\label{fig:agg}
\end{figure}

Since we aim to identify devices using any protocol, be it WiFi, Bluetooth, ZigBee or Z-Wave, we only use individual packets in the identification step, 
and use an \textit{aggregation algorithm} when identifying features such as MAC addresses are available. Thus, the identifying features are not used to create the models, but rather to better group similarly labelled packets to improve overall model performance.
The aggregation algorithm consists of two steps (see Fig.~\ref{fig:agg}), using as input the MAC address and the predicted label.  In the first step, it groups the MAC addresses according to the labels assigned to them and then finds the predominant MAC address for each label. If a MAC address is selected as dominant for more than one label, this MAC address is added to the exception list (likely a transfer problem with one MAC address being used for more than one device). 
In the second step, the predicted labels are gathered together in groups according to their MAC addresses. The most repetitive label among these groups is applied to the whole group to obtain aggregated labels. This procedure is not applied for MAC addresses that have entered the exception list; only the individual results are used for them. The device identification process assumes a benign network environment. Although the aggregation algorithm is effective for benign data, there is a risk of grouping together malicious packets that imitate legitimate IP/MAC addresses and display similar behaviour to benign packets with the same IP/MAC address. So, when applying the aggregation algorithm to networks with malicious data, caution is advised. 


In IoTDevID~\cite{kostas2022iot}, we used two datasets,
from Aalto  University~\cite{miettinen2017iot,aalto2017dataset} and UNSW ~\cite{sivanathan2018classifying}, which were produced for device identification studies. We used the Aalto dataset to develop our method and the UNSW dataset to validate our results. 
With the Aalto dataset (27 devices) IoTDevID achieved a 86.10\% F1 score, with 93.70\% for the UNSW dataset (32 devices).
Both datasets contain data generated from real device behaviour and have been used by most previous studies on device identification. 
%
%
However, they have limitations. The Aalto dataset consists only of packets captured during device setup, 
not
actual usage. 
The UNSW dataset contains data from different  sessions but lacks information about the nature of device use (\state{active} or \state{idle}). Additionally, it does not support the ability to aggregate devices under the same label (such as two different devices of the same brand and model) or observe non-IP devices (because the entire dataset consists only of IP devices), which limited the analysis of the transfer problem.

 In 2022, a new  dataset, CIC-IoT-22~\cite{CIC}, was made public.
It contains more devices than both Aalto and UNSW and its traffic was recorded whilst devices were operating in a wider range of activity states (e.g., \state{active} and \state{idle}). Hence it addresses some of the previous dataset issues. 
It also maintains the advantageous properties. 
For example, like the Aalto dataset, it has multiple instances of some devices and non-IP devices. Like the UNSW dataset, it contains long-term usage data.  In Section~\ref{Case}, 
we will validate the methodology used in the IoTDevID study by analyzing the CIC-IoT-22 dataset.


\section{Case Study} \label{Case} 

In this section, the CIC-IoT-22 dataset is examined and its features are analysed in depth. The individual packets and aggregation methods used for classification are explained. Finally, how feature extraction and labelling are performed is described.

\subsection{CIC-IoT-22 dataset} \label{data}

Data was collected during 6 different device states. These states can be summarised as follows~\cite{CIC}. 
In the \textbf{Power} state, each device is isolated from other devices and rebooted and the network packets related to this device are collected.
In the \textbf{Interactions} state, the device is interacted with by buttons, applications or voice commands and the network packets generated during this process are collected.
In \textbf{Scenarios}, the network data of these devices are collected in scenarios such as entering the house, leaving the house, unauthorised entry to the house at night and day or user error. 
In the \textbf{Attack} state, data is collected by applying Flood attacks and RTSP Brute Force attacks to the devices.
The \textbf{idle} state consists of recording every 8-hour period for 30 days in the evening hours when the devices are working but not actively used.
The \textbf{Active} state contains the data of the devices being used during the day for 30 days. This data is generated by people entering the lab and using the devices.

Some important points about the dataset:
In this study, the most important sections for benign device behaviour are \state{idle} and \state{active} as these states cover most normal usage and provide a wide range of data about all devices. Although it is stated in the paper~\cite{CIC} that 60 devices were used in this process, according to our experiments and the information provided in the dataset\footnote{\href{http://205.174.165.80/IOTDataset/CIC_IOT_Dataset2022/}{http://205.174.165.80/IOTDataset/CIC\_IOT\_Dataset2022}}, the data for these states covers 40 devices. These 40 devices are only LAN/Wired or WIFI devices, they do not include Zigbee and Z-Wave devices. Zigbee and Z-Wave devices have data isolated from other devices in the \state{power} and \state{interaction} stages. However, these data are both very limited and do not contain normal usage data. Also, the data of the Z-Wave devices is not in pcap format.


\subsection{Feature Extraction and Labelling}

\href{https://www.python.org/}{Python}, \href{https://scapy.net/}{Scapy}, and \href{https://www.wireshark.org/}{Wireshark} were used for feature extraction from  packet capture (pcap) files. Only individual packet-based features are used for feature extraction. Many of these features are derived from packet headers, but there are also payload-based features such as payload entropy and payload bytes. Although the feature extraction system created about 100 features\footnote{Complete feature list:\href{https://github.com/kahramankostas/IoTDevID-CIC/blob/main/featurelist.md}{github.com/kahramankostas/IoTDevID-CIC/blob/main/featurelist.md}} in total, 
only the features\footnote {Selected features are: pck\_size, Ether\_type, LLC\_ctrl, EAPOL\_version, EAPOL\_type, IP\_ihl, IP\_tos, IP\_len, IP\_flags, IP\_DF, IP\_ttl, IP\_options, ICMP\_code, TCP\_dataofs, TCP\_FIN, TCP\_ACK, TCP\_window, UDP\_len, DHCP\_options, BOOTP\_hlen, BOOTP\_flags, BOOTP\_sname, BOOTP\_file, BOOTP\_options, DNS\_qr, DNS\_rd, DNS\_qdcount, dport\_class, payload\_bytes, entropy} selected during the feature selection phase of the IoTDevID study were used in these experiments.

Labelling was performed using the list of device names/MAC address pairs in the dataset. In each fingerprint (feature set representing a packet) extracted, the source MAC address part was replaced with the given name. The MAC addresses not given in this list (5 MAC addresses that we believe belong to the hub, switch or the computer where the data is collected) were ignored.

In the CIC-IoT-22 dataset, each of the pcap files we use for feature extraction contains network traffic recorded on a day, and is named with the date it was recorded.   For example, data recorded on 24.11.2021 is labelled A211124 if \state{active} and I211124 if \state{idle}. In this context, 30 \state{idle} and 24~\footnote{Although the paper\cite{CIC} states 30 \state{active} sessions, there are only 24 sessions in the data set.} \state{active} sessions were recorded. As a preliminary study, we aimed to test the performance of models trained on data from each session by comparing them with each other. In order to compare the sessions with each other, they should contain the same devices. Unfortunately, data was not collected from every device in every session, and in some sessions, some devices did not generate any data at all. Table~\ref{tab:packets} shows how much data was generated by each device in each session in terms of network packets. Therefore, we only compare sessions that contain the same devices with each other. For this comparison, we create a session ID. In this ID, each device is represented by a binary digit. If the session has that device, it is indicated with 1, if not, it is indicated with 0. For example, if Session1 contains devices A and C, but not device B, then the ID number is 101(ABC). Session1 can be compared to other sessions with the same ID number without any problem. In this context, we have created a 32-digit ID for each session according to  32 device classes in total. There are 40 physical devices; however, some of these devices are in the same label because they are identical in brand and model.

\section{Results}\label{Evaluation}

We first consider data quality, in terms of its ability to support the training of device identification models, by training and testing ML models using different sessions within the CIC-IoT-2022 dataset. Based on the findings of this analysis, appropriate sessions are then merged to produce a dataset that is both representative of the data diversity and which can be reliably used to train device identification models. 

\subsection{Analysis of Data Quality}


We begin by training models using data from each session and then testing them on other sessions with the same ID. 1036 session pairs were created for this purpose, with the first session used for training and the second for testing. These pairs were divided into \state{active} and \state{idle} categories, resulting in four training vs testing possibilities: \state{active} vs \state{active} (AA), \state{active} vs \state{idle} (AI), \state{idle} vs \state{active} (IA), and \state{idle} vs \state{idle} (II). 
We utilize the F1 score as a primary metric for reporting results for two key reasons. First, unlike accuracy, the F1 score offers reliable performance evaluation on unbalanced datasets, which is often the case with IoT-generated data, including this study. Second, the F1 score provides insights not only into overall performance but also class-specific performance, enabling detailed analysis. However, for the sake of comprehensive assessment, we also include accuracy as a comparative measure, given its prevalent (if sometimes inappropriate) usage in the literature.

Table~\ref{tab:PP} presents the average results for 1036 session pairs categorized into four conditions. When we focus on individual results, the highest F1 score is achieved in condition II (72\%), closely followed by AI (71.4\%). The lowest scores are observed in the AA (70.2\%) and IA (67.9\%) conditions, respectively.  When we apply the packet aggregation algorithm, it can be seen that the results reflect those from individual packets, but with an improvement of approximately 5-7 points in each case.
Notably, the utilization of \state{idle} data for testing purposes yields higher performance compared to the use of \state{active} data. This can be attributed to the broader range of data available in \state{active} scenarios, while \state{idle} data lacks this diversity. Consequently, a model trained with \state{active} data exhibits higher success when tested on \state{idle} data, whereas a model trained with \state{idle} data shows lower performance when tested on \state{active} data.


\begin{table}[htbp]
  \centering
\caption{Results for four device conditions using Decision Tree (DT). Results consisting of accuracy and F1 scores (with standard deviations) are the means of all session comparisons. t specifies the time in seconds.}
\setlength{\tabcolsep}{6pt}
    \begin{tabular}{@{}clllrrr@{}}
    \toprule
          & Data  & Accuracy & F1-Score & \multicolumn{1}{l}{Train-t} & \multicolumn{1}{l}{Test-t} & \multicolumn{1}{l}{Ag-t} \\
    \midrule
    \multirow{4}[2]{*}{\begin{sideways}Individual\end{sideways}} & AA    
& 0.756±0.001 & 0.702±0.003 & 1.620  & 0.185 & 0 \\
          & AI    & 0.760±0.002 & 0.714±0.004 & 1.528 & 0.191 & 0 \\
          & IA    & 0.686±0.002 & 0.679±0.004 & 1.545 & 0.171 & 0 \\
          & II    & \underline{0.768±0.003} & \underline{0.720±0.005} & 1.516 & 0.194 & 0 \\
    \midrule
    \multirow{4}[2]{*}{\begin{sideways}Aggregated\end{sideways}} & AA    
& 0.810±0.001 & 0.773±0.004 & 1.570  & 0.184 & 7.487 \\
          & AI    & 0.815±0.002 & \underline{0.780±0.005} & 1.491 & 0.190  & 7.475 \\
          & IA    & 0.731±0.002 & 0.744±0.005 & 1.620  & 0.173 & 7.285 \\
          & II    & \underline{0.819±0.004} & 0.778±0.006 & 1.470  & 0.175 & 7.015 \\
    \bottomrule
    \end{tabular}%
  \label{tab:PP}%
\end{table}%

\begin{figure}[ht]

\subfloat[17X17 F1 Score list\label{fig:comp_2}\centering ]{{\includegraphics[width=65mm]{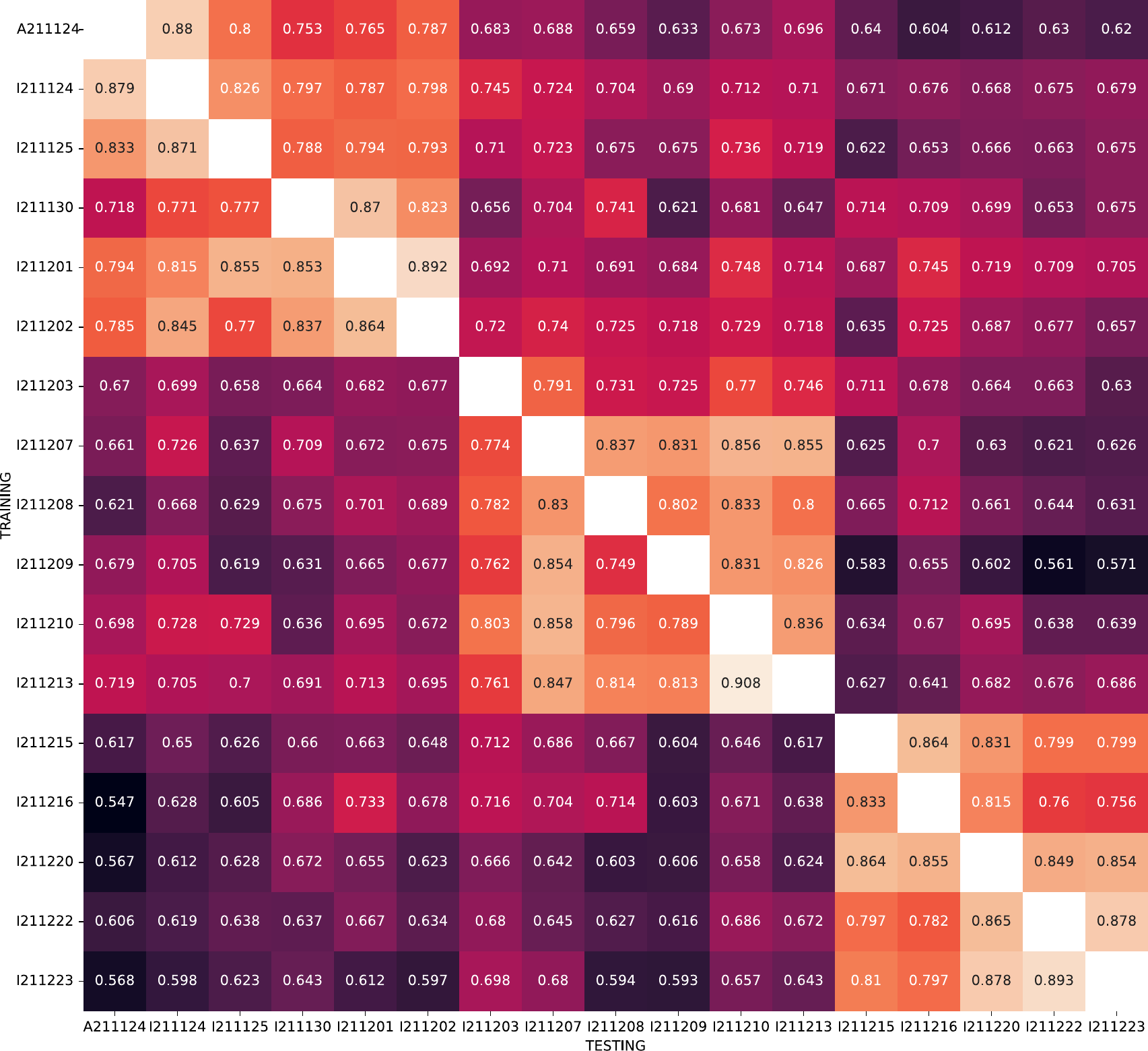}}}%
\subfloat [28X28 F1 Score list\label{fig:comp_1}\centering ]{{\includegraphics[width=65mm]{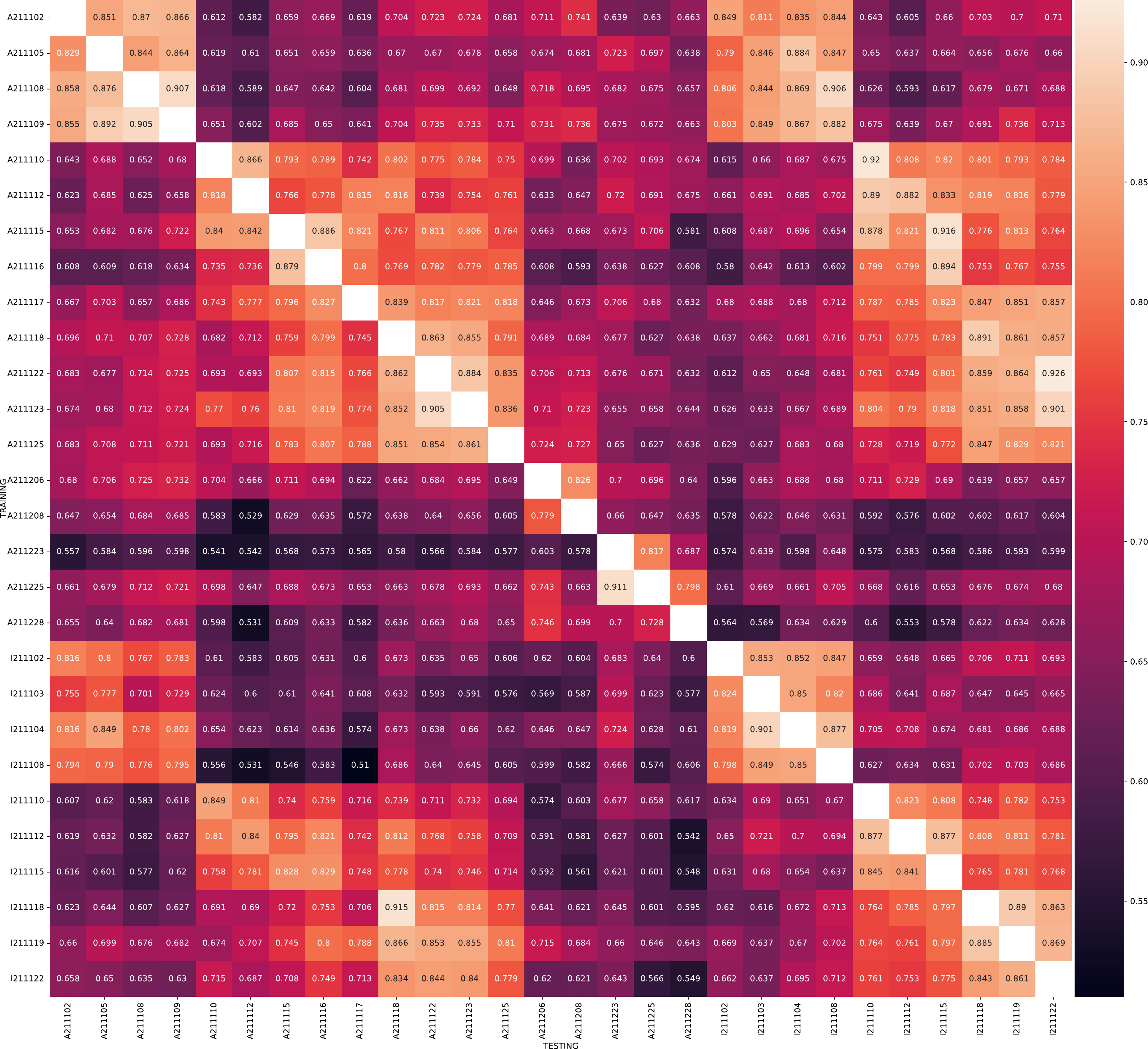}}}
\caption{Comparison of sessions with the dataset's same session ID (device distribution). Y-axis shows training and X-axis shows testing data. These results in F1 scores are obtained using DT.}
	\label{fig:stepsofID}
\end{figure}

To gain a more comprehensive understanding of the data, it is important to analyze individual cases. Fig.~\ref{fig:stepsofID} shows a heatmap displaying the F1 scores obtained from session pairs. The vertical axis represents the training data, while the horizontal axis represents the test data. The F1 score ranges from 51\% to 93\% in pairwise session comparisons. It is important to note that this is a multiple-classification task with approximately 32 classes. In contrast to binary classification, where results above 50\% are considered significant, a randomly assigned multiple-classification model would achieve an accuracy of approximately 3.1\% (100 divided by 32). Therefore, even an F1 score of 51\% represents a substantial improvement over chance/random success.


In Fig.~\ref{fig:comp_2} \&~\subref*{fig:comp_1} the session IDs that allow the broadest comparison, containing 28 and 17 sessions respectively, are shown. Fig.~\ref{fig:comp_2} predominantly consists of \state{idle} examples, showing higher success rates when comparing consecutive dates. 
The heatmap exhibits a somewhat symmetrical structure, albeit imperfect, particularly due to minimal user intervention during \state{idle} collection. In  Fig.~\ref{fig:comp_1}, both \state{active} and \state{idle} sessions are mixed. Similarly, success rates are higher for consecutive sessions. However, the involvement of users introduces a more distorted symmetry, especially in \state{active} sessions. Significant performance drops are observed when using data collected on specific dates coinciding with a national holiday, such as 
2021-12-23, 2021-12-25, and 2021-12-28. Additionally, the lowest performances occur when using \state{idle} as training data and \state{active} as test data. We believe that \state{active} sessions offer a broader representation, because \state{active} use actually includes \state{idle} use as well, while it is not possible to say the opposite. However, the inherent differences in data collection during \state{active} sessions change the model's performance. So it is not possible to speak of perfect patterns when human factors are involved. 
The subsequent section explores whether
increasing the diversity by
combining data from different sessions improves representation and model performance.

\subsection{Dataset Construction}\label{moredata}

\begin{figure}[ht]
	\centerline{\includegraphics[width=.70\columnwidth]{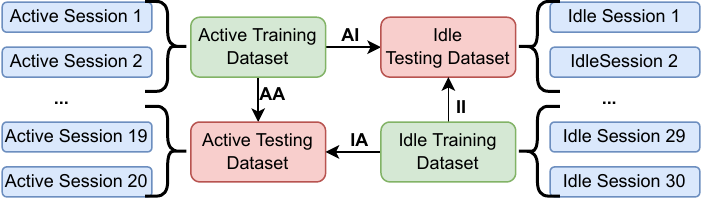}}
	\caption{Creating datasets with larger representation by combining sessions. Training and testing data of both types (\state{active} and \state{idle}) were obtained and they were used in the training and testing of models both for the same category of data (\state{active} training vs \state{active} Testing) and for their complement (\state{active} Training vs \state{idle} Testing).}
	\label{fig:data}
\end{figure}

We aimed to enhance sample diversity and improve model performance by sampling from multiple sessions. The data already included \state{idle} and \state{active} sessions, which we further split into training and testing subsets. This resulted in four subsets: \state{idle}-training, \state{idle}-testing, \state{active}-training, and \state{active}-testing, derived from a total of 54 sessions. Refer to Table~\ref{tab:packets} (Appendix) for the specific assignment of sessions to each subset. The dataset creation process is illustrated in Fig.~\ref{fig:data}.


However, due to some deficiencies in the dataset, we have made minor changes to the data. We copied the data of the D-Link Water Sensor, a device not included in the \state{active} sessions, from the \state{idle} sessions to the active session data. Another change was related to the LG Smart TV device. The data for this device is only present in three of the 54 sessions. Furthermore, the data for this device is so unbalanced that the data in these three sessions account for about 9\% of the total number of packets in all 40 devices. For these reasons, we removed this device from the dataset.

To ensure a balanced dataset that represents session diversity without excessive size, we reduced the number of packets in each of the four datasets to 10\% of the total number of packets per dataset by random sampling. This random sampling was employed during this process to maintain consistent packet rates for each device, preserving the natural distribution of the dataset. 

\subsection{IoTDevID Evaluation}\label{moredataresult}

Next, we evaluate the performance of the IoTDevID methodology on the datasets described above. Table~\ref{tab:result} summarises the test performances, both when using individual packets and when using the aggregation algorithm. It can be seen that very good results are obtained in all cases, with all F1 scores being above 81\%. When individual packets are used, it is seen that the most successful model is AI with 90.50\%, followed by AA with 84.20\%, while the results of cases IA and II are very close to each other with a score of around 81\%.
As in the original study, a significant further improvement is seen when the aggregation algorithm is used. The AA and IA cases improve by about eight points, and the AI and II cases improve by about 10 points. For AI, the models achieve almost perfect discrimination.

\begin{table}[htbp]
	\centering
	\caption{Results for four device conditions with individual and aggregated methods using DT. Results consisting of accuracy and F1 scores (with standard deviations) are the means of 100 repeats. t specifies time in seconds. Ag-t is the duration of the aggregation algorithm. }
 \setlength{\tabcolsep}{6pt}
	\begin{tabular}{@{}clllrrr@{}}
		\toprule
		& Data  & Accuracy & F1 Score & \multicolumn{1}{l}{Train-t} & \multicolumn{1}{l}{Test-t} & \multicolumn{1}{l}{Ag-t} \\
		\midrule
		\multirow{4}[2]{*}{\begin{sideways}Individual\end{sideways}} & AA    & 0.890±0.001 & 0.842±0.004 & 1.748 & 0.204 & 0 \\
		& AI    & \underline{0.918±0.001}  & \underline{0.905±0.005} & 1.812 & 0.287 & 0 \\
		& IA    & 0.823±0.046 & 0.818±0.015 & 1.699 & 0.223 & 0 \\
		& II    & 0.821±0.004 & 0.814±0.007 & 1.721 & 0.291 & 0 \\
		\midrule
		\multirow{4}[2]{*}{\begin{sideways}Aggregated\end{sideways}} & AA    & 0.943±0.001 & 0.925±0.007 & 1.962 & 0.235 & 9.119 \\
		& AI    & \underline{0.999±0.000} & \underline{0.999±0.000} & 1.864 & 0.299 & 11.519 \\
		& IA    & 0.850±0.058 & 0.898±0.017 & 1.584 & 0.206 & 8.46 \\
		& II    & 0.904±0.004 & 0.912±0.006 & 1.630  & 0.313 & 11.267 \\
		\bottomrule
	\end{tabular}%
	\label{tab:result}%
\end{table}%

Upon comparing these results with Table~\ref{tab:PP}, it is evident that all scores have exhibited significant improvements. Analyzing individual results, the F1 score has risen from 70.2 to 84.2 for AA, from 71.4 to 90.4 for AI, from 67.9 to 81.8 for IA, and from 72 to 81.4 for II. Notably, the aggregation results demonstrate an even greater increase. The choice of data used exerts a substantial influence on the model's performance, underscoring the importance of constructing a more representative dataset through data combination. This enhanced dataset has substantially bolstered the success of models trained using it.

Returning to Table~\ref{tab:result}, the \state{active} state has a broader representation as it includes network data both when the devices are used and not used. \state{Idle} includes only passive states and does not include the states when the devices are used. The AI case, which employs the \state{idle} dataset as testing, exhibits an exceptionally high performance that may not reflect practical conditions, as the uniform data distribution of the \state{idle} dataset creates an ``easier'' testing environment. In this context, using the \state{active} state for both training and testing gives more realistic results.

Further analysis at the class level allows for a deeper understanding. In this context, Table~\ref{tab:class} shows the class-based F1 scores of all devices and Fig.~\ref{fig:cmson} shows the confusion matrix for the AA case. Focusing on the aggregated results in column AA, it is evident that 22 out of 31 devices achieve near-perfect classification with an F1 score above
99\%. Six devices (Globe Lamp, Gosund Plug, HeimVision Lamp, Teckin Plug, Yutron Plug) achieve lower performances, although still above 90\%. These devices are lamps or plugs serving similar functions. The dataset is also rich in cameras and speakers, forming another group of devices with similar tasks. However, the classification of these devices does not pose similar challenges. This can be attributed to the fact that devices such as lamps and sockets have simpler structures compared to speakers and cameras, leading to similar data outputs that are more difficult to discriminate. Similar difficulties were encountered in previous experiments with sensors, plugs and switches in the IoTDevID study using the Aalto dataset\cite{kostas2022iot}.

Noteworthy are the devices Ring Base Station, Amazon AE Spot and Smart Board, which exhibit significantly lower F1 scores than the other devices. Packets from Ring Base Station devices are often misclassified as speakers (Amazon Alexa family, Sonos Speaker, etc.), likely due to their role as a link between alarm systems in smart homes and management systems like Alexa or other speakers.
Analyzing the results for the Amazon AE Spot device, although the recall is high, the precision is exceptionally low (refer to Table~\ref{tab:class-aa} in Appendix).
The Smart Board device poses a challenge as a majority of packets are mislabeled as Amazon AE Spot. Examining the data distribution of the Smart Board, an outlier is observed in the A211126 data, which was added to the \state{active} test dataset (see Table~\ref{tab:packets}  in Appendix). On this specific day, the data collected for this device is twice the combined amount of the other 53 days. Moreover, 78.6\% of this unusually large data volume, which is 100 times greater than other sessions, 
comprises uniformly empty packets that are challenging to classify (TCP packets with the ACK flag set and no payload). Although our study does not analyze the causes of these outliers, it is important to note that the imbalanced data distribution resulting from this outlier greatly complicates the identification of this device during the test phase.

\begin{table}[htbp]

  \centering
	\setlength{\tabcolsep}{2pt}
	\caption{Class-based results for four device conditions with individual and aggregated methods using DT. Results consisting of F1 scores are the means of 100 repeats.}
	\begin{tabular}{@{}lrrrr|rrrr@{}}

\cmidrule{2-9}          & \multicolumn{4}{c}{Individual} & \multicolumn{4}{c}{ Aggregated} \\
       \hline
          & \multicolumn{1}{c}{AA} & \multicolumn{1}{c}{AI} & \multicolumn{1}{c}{IA} & \multicolumn{1}{c|}{II} & \multicolumn{1}{c}{AA} & \multicolumn{1}{c}{AI} & \multicolumn{1}{c}{IA} & \multicolumn{1}{c}{II} \\
    \hline
    Amcrest WiFi-Cam. & \cellcolor[rgb]{ 1,  .965,  .965}0.968 & \cellcolor[rgb]{ 1,  .976,  .976}0.979 & \cellcolor[rgb]{ 1,  .945,  .945}0.951 & \cellcolor[rgb]{ 1,  .953,  .953}0.959 & \cellcolor[rgb]{ 1,  .988,  .988}0.992 & 1.000 & \cellcolor[rgb]{ 1,  .996,  .996}1.000 & \cellcolor[rgb]{ 1,  .996,  .996}1.000 \\
    Amazon AE Dot & \cellcolor[rgb]{ 1,  .925,  .925}0.933 & \cellcolor[rgb]{ 1,  .929,  .929}0.938 & \cellcolor[rgb]{ 1,  .945,  .945}0.950 & \cellcolor[rgb]{ 1,  .941,  .941}0.947 & \cellcolor[rgb]{ 1,  .996,  .996}0.997 & \cellcolor[rgb]{ 1,  .996,  .996}1.000 & \cellcolor[rgb]{ 1,  .996,  .996}0.998 & \cellcolor[rgb]{ 1,  .996,  .996}1.000 \\
    Amazon AE Spot & \cellcolor[rgb]{ 1,  .518,  .518}0.555 & \cellcolor[rgb]{ 1,  .824,  .824}0.838 & \cellcolor[rgb]{ 1,  .827,  .827}0.844 & \cellcolor[rgb]{ 1,  .824,  .824}0.837 & \cellcolor[rgb]{ 1,  .522,  .522}0.559 & \cellcolor[rgb]{ 1,  .996,  .996}1.000 & \cellcolor[rgb]{ 1,  .996,  .996}0.999 & \cellcolor[rgb]{ 1,  .996,  .996}0.999 \\
    Amazon AE Studio & \cellcolor[rgb]{ 1,  .804,  .804}0.821 & \cellcolor[rgb]{ 1,  .863,  .863}0.874 & \cellcolor[rgb]{ 1,  .824,  .824}0.837 & \cellcolor[rgb]{ 1,  .714,  .714}0.736 & \cellcolor[rgb]{ 1,  .976,  .976}0.981 & \cellcolor[rgb]{ 1,  .996,  .996}1.000 & \cellcolor[rgb]{ 1,  .996,  .996}0.999 & \cellcolor[rgb]{ 1,  .976,  .976}0.979 \\
    Amazon Plug & \cellcolor[rgb]{ 1,  .992,  .992}0.995 & \cellcolor[rgb]{ 1,  .996,  .996}0.999 & \cellcolor[rgb]{ 1,  .996,  .996}0.997 & \cellcolor[rgb]{ 1,  .996,  .996}0.999 & \cellcolor[rgb]{ 1,  .996,  .996}1.000 & 1.000 & \cellcolor[rgb]{ 1,  .996,  .996}1.000 & 1.000 \\
    Arlo Base Station & \cellcolor[rgb]{ 1,  .98,  .98}0.984 & \cellcolor[rgb]{ 1,  .804,  .804}0.819 & \cellcolor[rgb]{ 1,  .592,  .592}0.624 & \cellcolor[rgb]{ 1,  .851,  .851}0.864 & 1.000 & \cellcolor[rgb]{ 1,  .996,  .996}0.998 & \cellcolor[rgb]{ 1,  .408,  .408}0.454 & \cellcolor[rgb]{ 1,  .996,  .996}1.000 \\
    Arlo Q Camera & \cellcolor[rgb]{ 1,  .984,  .984}0.987 & \cellcolor[rgb]{ 1,  .965,  .965}0.970 & \cellcolor[rgb]{ 1,  .965,  .965}0.969 & \cellcolor[rgb]{ 1,  .945,  .945}0.952 & 1.000 & 1.000 & \cellcolor[rgb]{ 1,  .996,  .996}1.000 & \cellcolor[rgb]{ 1,  .996,  .996}1.000 \\
    Atomi Coff-Maker & \cellcolor[rgb]{ 1,  .831,  .831}0.847 & \cellcolor[rgb]{ 1,  .882,  .882}0.892 & \cellcolor[rgb]{ 1,  .878,  .878}0.890 & \cellcolor[rgb]{ 1,  .459,  .459}0.501 & \cellcolor[rgb]{ 1,  .996,  .996}0.999 & \cellcolor[rgb]{ 1,  .996,  .996}1.000 & 1.000 & \cellcolor[rgb]{ 1,  .608,  .608}0.638 \\
    Borun Camera & \cellcolor[rgb]{ 1,  .976,  .976}0.982 & \cellcolor[rgb]{ 1,  .976,  .976}0.981 & \cellcolor[rgb]{ 1,  .969,  .969}0.972 & \cellcolor[rgb]{ 1,  .973,  .973}0.978 & \cellcolor[rgb]{ 1,  .996,  .996}0.999 & \cellcolor[rgb]{ 1,  .996,  .996}0.999 & \cellcolor[rgb]{ 1,  .996,  .996}0.999 & \cellcolor[rgb]{ 1,  .996,  .996}0.999 \\
    D-Link Mini Cam. & \cellcolor[rgb]{ 1,  .98,  .98}0.982 & \cellcolor[rgb]{ 1,  .984,  .984}0.989 & \cellcolor[rgb]{ 1,  .286,  .286}0.342 & \cellcolor[rgb]{ 1,  .898,  .898}0.906 & 1.000 & \cellcolor[rgb]{ 1,  .996,  .996}1.000 & \cellcolor[rgb]{ 1,  .733,  .733}0.756 & \cellcolor[rgb]{ 1,  .996,  .996}1.000 \\
    D-Link Water Sen. & \cellcolor[rgb]{ 1,  .922,  .922}0.930 & \cellcolor[rgb]{ 1,  .929,  .929}0.936 & \cellcolor[rgb]{ 1,  .929,  .929}0.936 & \cellcolor[rgb]{ 1,  .925,  .925}0.934 & 1.000 & \cellcolor[rgb]{ 1,  .984,  .984}0.989 & 1.000 & \cellcolor[rgb]{ 1,  .984,  .984}0.989 \\
    Eufy HomeBase 2 & \cellcolor[rgb]{ 1,  .8,  .8}0.815 & \cellcolor[rgb]{ 1,  .796,  .796}0.812 & \cellcolor[rgb]{ 1,  .761,  .761}0.782 & \cellcolor[rgb]{ 1,  .788,  .788}0.806 & \cellcolor[rgb]{ 1,  .996,  .996}1.000 & \cellcolor[rgb]{ 1,  .996,  .996}0.998 & \cellcolor[rgb]{ 1,  .996,  .996}1.000 & \cellcolor[rgb]{ 1,  .996,  .996}0.998 \\
    Globe Lamp  & \cellcolor[rgb]{ 1,  .624,  .624}0.654 & \cellcolor[rgb]{ 1,  .808,  .808}0.823 & \cellcolor[rgb]{ 1,  .906,  .906}0.916 & \cellcolor[rgb]{ 1,  .384,  .384}0.431 & \cellcolor[rgb]{ 1,  .89,  .89}0.900 & \cellcolor[rgb]{ 1,  .996,  .996}0.997 & 1.000 & \cellcolor[rgb]{ 1,  .184,  .184}0.246 \\
    Google Nest Mini & \cellcolor[rgb]{ 1,  .973,  .973}0.975 & \cellcolor[rgb]{ 1,  .969,  .969}0.971 & \cellcolor[rgb]{ 1,  .945,  .945}0.952 & \cellcolor[rgb]{ 1,  .867,  .867}0.879 & \cellcolor[rgb]{ 1,  .996,  .996}1.000 & \cellcolor[rgb]{ 1,  .996,  .996}1.000 & \cellcolor[rgb]{ 1,  .992,  .992}0.996 & \cellcolor[rgb]{ 1,  .976,  .976}0.982 \\
    Gosund Plug & \cellcolor[rgb]{ 1,  .824,  .824}0.839 & \cellcolor[rgb]{ 1,  .89,  .89}0.899 & \cellcolor[rgb]{ 1,  .91,  .91}0.917 & \cellcolor[rgb]{ 1,  .682,  .682}0.706 & \cellcolor[rgb]{ 1,  .965,  .965}0.968 & \cellcolor[rgb]{ 1,  .992,  .992}0.995 & \cellcolor[rgb]{ 1,  .996,  .996}0.999 & \cellcolor[rgb]{ 1,  .702,  .702}0.724 \\
    Gosund Socket & \cellcolor[rgb]{ 1,  .855,  .855}0.868 & \cellcolor[rgb]{ 1,  .882,  .882}0.893 & \cellcolor[rgb]{ 1,  .882,  .882}0.895 & \cellcolor[rgb]{ 1,  .494,  .494}0.532 & \cellcolor[rgb]{ 1,  .988,  .988}0.992 & \cellcolor[rgb]{ 1,  .992,  .992}0.995 & \cellcolor[rgb]{ 1,  .996,  .996}0.999 & \cellcolor[rgb]{ 1,  .357,  .357}0.406 \\
    HeimVision S Cam. & \cellcolor[rgb]{ 1,  .984,  .984}0.986 & \cellcolor[rgb]{ 1,  .996,  .996}0.998 & \cellcolor[rgb]{ 1,  .925,  .925}0.934 & \cellcolor[rgb]{ 1,  .98,  .98}0.984 & 1.000 & \cellcolor[rgb]{ 1,  .996,  .996}1.000 & 1.000 & \cellcolor[rgb]{ 1,  .996,  .996}1.000 \\
    HeimVision Lamp & \cellcolor[rgb]{ 1,  .69,  .69}0.715 & \cellcolor[rgb]{ 1,  .824,  .824}0.838 & \cellcolor[rgb]{ 1,  .843,  .843}0.857 & \cellcolor[rgb]{ 1,  .478,  .478}0.518 & \cellcolor[rgb]{ 1,  .961,  .961}0.965 & \cellcolor[rgb]{ 1,  .996,  .996}0.999 & 1.000 & \cellcolor[rgb]{ 1,  .737,  .737}0.759 \\
    Home Eye Camera & \cellcolor[rgb]{ 1,  .922,  .922}0.930 & \cellcolor[rgb]{ 1,  .902,  .902}0.911 & \cellcolor[rgb]{ 1,  .918,  .918}0.927 & \cellcolor[rgb]{ 1,  .902,  .902}0.910 & \cellcolor[rgb]{ 1,  .996,  .996}1.000 & \cellcolor[rgb]{ 1,  .996,  .996}1.000 & \cellcolor[rgb]{ 1,  .996,  .996}1.000 & \cellcolor[rgb]{ 1,  .996,  .996}1.000 \\
    Luohe Cam Dog & \cellcolor[rgb]{ 1,  .741,  .741}0.762 & \cellcolor[rgb]{ 1,  .737,  .737}0.760 & \cellcolor[rgb]{ 1,  .737,  .737}0.759 & \cellcolor[rgb]{ 1,  .733,  .733}0.757 & \cellcolor[rgb]{ 1,  .996,  .996}1.000 & \cellcolor[rgb]{ 1,  .992,  .992}0.995 & \cellcolor[rgb]{ 1,  .996,  .996}1.000 & \cellcolor[rgb]{ 1,  .992,  .992}0.994 \\
    Nest Indoor Cam. & \cellcolor[rgb]{ 1,  .996,  .996}0.998 & \cellcolor[rgb]{ 1,  .996,  .996}0.997 & \cellcolor[rgb]{ 1,  .996,  .996}0.999 & \cellcolor[rgb]{ 1,  .902,  .902}0.910 & \cellcolor[rgb]{ 1,  .996,  .996}0.999 & \cellcolor[rgb]{ 1,  .996,  .996}0.999 & \cellcolor[rgb]{ 1,  .996,  .996}1.000 & \cellcolor[rgb]{ 1,  .996,  .996}0.997 \\
    Netatmo Camera & \cellcolor[rgb]{ 1,  .965,  .965}0.969 & \cellcolor[rgb]{ 1,  .984,  .984}0.986 & \cellcolor[rgb]{ 1,  .349,  .349}0.398 & \cellcolor[rgb]{ 1,  .925,  .925}0.934 & \cellcolor[rgb]{ 1,  .996,  .996}1.000 & \cellcolor[rgb]{ 1,  .996,  .996}1.000 & \cellcolor[rgb]{ 1,  .329,  .329}0.381 & \cellcolor[rgb]{ 1,  .996,  .996}1.000 \\
    Netatmo Weather & \cellcolor[rgb]{ 1,  .808,  .808}0.826 & \cellcolor[rgb]{ 1,  .812,  .812}0.827 & \cellcolor[rgb]{ 1,  .855,  .855}0.868 & \cellcolor[rgb]{ 1,  .831,  .831}0.845 & 1.000 & \cellcolor[rgb]{ 1,  .996,  .996}1.000 & 1.000 & \cellcolor[rgb]{ 1,  .996,  .996}1.000 \\
    Philips Hue Bridge & \cellcolor[rgb]{ 1,  .992,  .992}0.994 & \cellcolor[rgb]{ 1,  .988,  .988}0.990 & \cellcolor[rgb]{ 1,  .973,  .973}0.978 & \cellcolor[rgb]{ 1,  .984,  .984}0.986 & 1.000 & \cellcolor[rgb]{ 1,  .996,  .996}1.000 & \cellcolor[rgb]{ 1,  .996,  .996}1.000 & 1.000 \\
    Ring Base Station & \cellcolor[rgb]{ 1,  .247,  .247}0.305 & \cellcolor[rgb]{ 1,  .902,  .902}0.913 & \cellcolor[rgb]{ 1,  .235,  .235}0.293 & \cellcolor[rgb]{ 1,  .671,  .671}0.697 & \cellcolor[rgb]{ 1,  .282,  .282}0.336 & \cellcolor[rgb]{ 1,  .996,  .996}1.000 & \cellcolor[rgb]{ 1,  .188,  .188}0.250 & \cellcolor[rgb]{ 1,  .992,  .992}0.995 \\
    SIMCAM 1S & \cellcolor[rgb]{ 1,  .996,  .996}0.996 & \cellcolor[rgb]{ 1,  .996,  .996}0.998 & \cellcolor[rgb]{ 1,  .976,  .976}0.980 & \cellcolor[rgb]{ 1,  .996,  .996}0.998 & \cellcolor[rgb]{ 1,  .996,  .996}1.000 & 1.000 & 1.000 & 1.000 \\
    Smart Board & \cellcolor[rgb]{ 1,  .31,  .31}0.363 & \cellcolor[rgb]{ 1,  .698,  .698}0.721 & \cellcolor[rgb]{ 1,  .235,  .235}0.292 & \cellcolor[rgb]{ 1,  .647,  .647}0.675 & \cellcolor[rgb]{ 1,  .031,  .031}0.105 & \cellcolor[rgb]{ 1,  .992,  .992}0.996 & \cellcolor[rgb]{ 1,  0,  0}0.074 & \cellcolor[rgb]{ 1,  .992,  .992}0.996 \\
    Sonos One Speaker & \cellcolor[rgb]{ 1,  .706,  .706}0.729 & \cellcolor[rgb]{ 1,  .831,  .831}0.844 & \cellcolor[rgb]{ 1,  .702,  .702}0.727 & \cellcolor[rgb]{ 1,  .882,  .882}0.891 & \cellcolor[rgb]{ 1,  .988,  .988}0.991 & \cellcolor[rgb]{ 1,  .996,  .996}0.999 & \cellcolor[rgb]{ 1,  .945,  .945}0.951 & \cellcolor[rgb]{ 1,  .996,  .996}1.000 \\
    Teckin Plug & \cellcolor[rgb]{ 1,  .647,  .647}0.675 & \cellcolor[rgb]{ 1,  .812,  .812}0.826 & \cellcolor[rgb]{ 1,  .855,  .855}0.868 & \cellcolor[rgb]{ 1,  .541,  .541}0.578 & \cellcolor[rgb]{ 1,  .925,  .925}0.932 & \cellcolor[rgb]{ 1,  .996,  .996}1.000 & \cellcolor[rgb]{ 1,  .996,  .996}1.000 & \cellcolor[rgb]{ 1,  .686,  .686}0.712 \\
    Yutron Plug & \cellcolor[rgb]{ 1,  .745,  .745}0.764 & \cellcolor[rgb]{ 1,  .843,  .843}0.855 & \cellcolor[rgb]{ 1,  .855,  .855}0.867 & \cellcolor[rgb]{ 1,  .6,  .6}0.632 & \cellcolor[rgb]{ 1,  .949,  .949}0.956 & \cellcolor[rgb]{ 1,  .996,  .996}1.000 & 1.000 & \cellcolor[rgb]{ 1,  .835,  .835}0.850 \\
    iRobot Roomba & \cellcolor[rgb]{ 1,  .949,  .949}0.955 & \cellcolor[rgb]{ 1,  .957,  .957}0.962 & \cellcolor[rgb]{ 1,  .827,  .827}0.843 & \cellcolor[rgb]{ 1,  .941,  .941}0.946 & 1.000 & 1.000 & \cellcolor[rgb]{ 1,  .988,  .988}0.993 & \cellcolor[rgb]{ 1,  .996,  .996}1.000 \\
    \hline
    Mean  & \cellcolor[rgb]{ 1,  .827,  .827}0.842 & \cellcolor[rgb]{ 1,  .894,  .894}0.905 & \cellcolor[rgb]{ 1,  .8,  .8}0.818 & \cellcolor[rgb]{ 1,  .796,  .796}0.814 & \cellcolor[rgb]{ 1,  .918,  .918}0.925 & \cellcolor[rgb]{ 1,  .996,  .996}0.999 & \cellcolor[rgb]{ 1,  .886,  .886}0.898 & \cellcolor[rgb]{ 1,  .902,  .902}0.912 \\
    \bottomrule
    \end{tabular}%
  \label{tab:class}%
\end{table}%

\begin{figure}[ht]
	\centerline{\includegraphics[width=1\columnwidth]{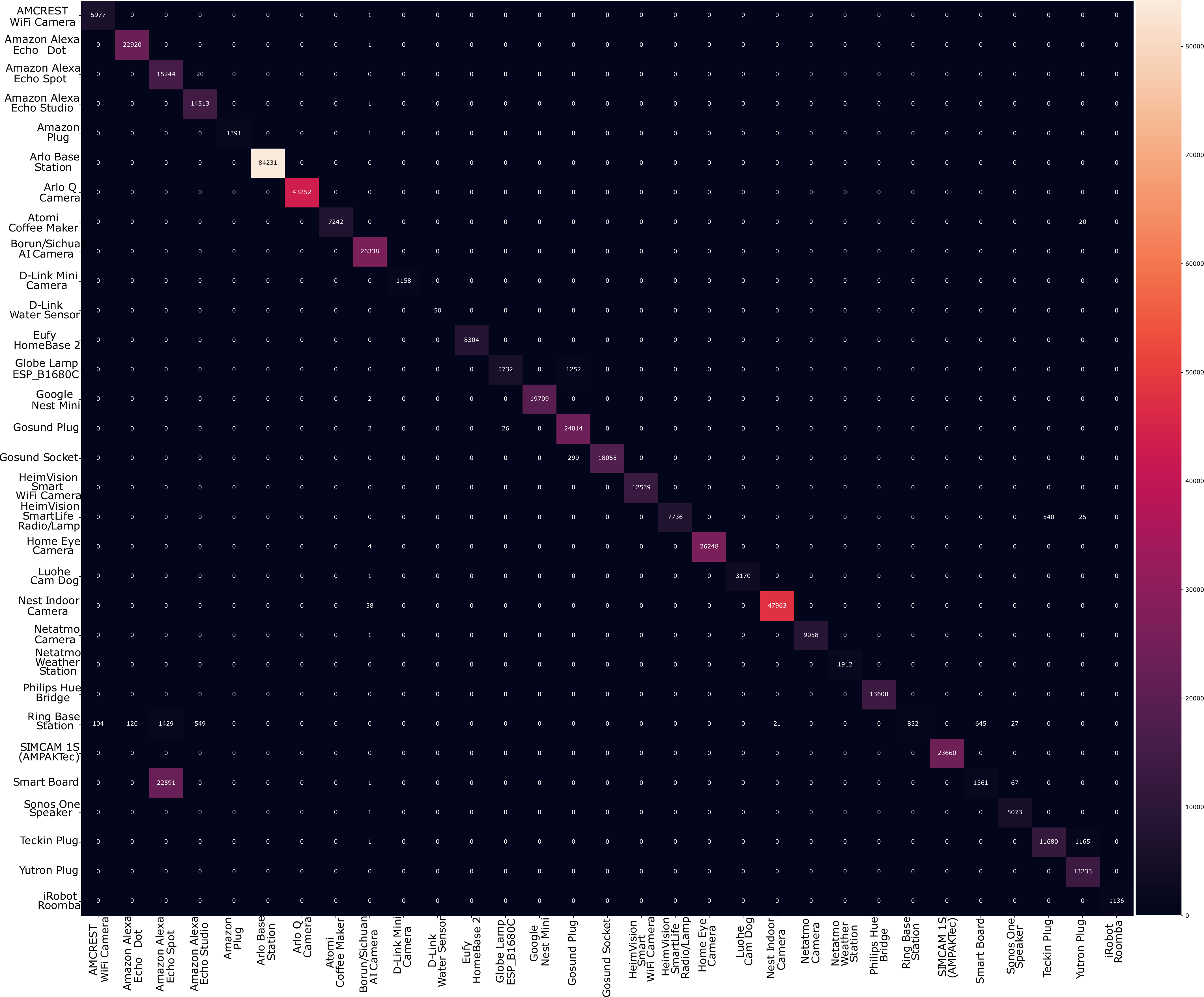}}
	\caption{Confusion matrix for case AA, showing means of 100 results using DT.}
	\label{fig:cmson}
\end{figure}

\subsection{Evaluation by including non-IP devices} \label{Including non-IP devices}

The CIC-IoT-2022 dataset also includes non-IP device (Zigbee and Z-Wave) data, specifically in the Power and Interactions states. Only Zigbee devices have records in raw network data format (pcap); Z-Wave device data is not available in this format, limiting the opportunity for feature extraction. Additionally, the non-IP devices lack normal usage data, and the amount of data collected in the Power and Interactions states is relatively small in comparison to IP devices. Table~\ref{tab:non-ip-packets} shows the number of packets collected in both cases. Despite these limitations, we found the non-IP device data in the experimental suite to be interesting for analysis.

Although the collection method of Zigbee device data does not perfectly align with the transfer problem case, it exhibits similarities due to the shared fixed MAC address (00:00:00:00:00:00) assigned to all devices. This provides an opportunity to explore the exception part of the aggregation algorithm. We incorporated the Zigbee data into the AA state by adding the Zigbee devices from the Interactions state (which had more data) to the training set and using the data from the Power state for testing. The class-based results can be found in Table~\ref{tab:non-ip-results}.
\begin{table}[htbp]
  \centering
  \caption{The total number of packets generated by the Zigbee devices in Interaction and Power conditions.}

    \begin{tabular}{@{}lccc|rlccl@{}}
\cline{1-3}\cline{6-8}          & \multicolumn{2}{c}{Number of Packets} &       &       &       & \multicolumn{2}{c}{Number of Packets} \\
\cline{2-3}\cline{7-8}    Device  & \multicolumn{1}{l}{Power} & \multicolumn{1}{l}{ Interactions} &       &       & Device  & \multicolumn{1}{l}{Power} & \multicolumn{1}{l}{ Interactions} \\
\cline{1-3}\cline{6-8}    AeoTec Button & 180   & 13    &       &       & Sengled Smart Plug & 367   & 203 \\
    AeoTec Motion Sensor & 173   & 25    &       &       & SmartThings Button & 166   & 12 \\
    AeoTec Multipurpose Sensor & 297   & 173   &       &       & SmartThings Hub & 198   & 0 \\
    AeoTec Water Leak Sensor & 340   & 24    &       &       & SmartThings  Bulb & 2870  & 1412 \\
    Philips Hue White & 2093  & 746   &       &       & Sonoff Smart Plug & 148   & 142 \\
\cline{1-3}\cline{6-8}    \end{tabular}%
  \label{tab:non-ip-packets}%
\end{table}%

\begin{table}[htbp]
  \centering
  \setlength{\tabcolsep}{3pt}
  \caption{Class-based results of IP and non-IP devices with AA case using individual and aggregated methods.}
    \begin{tabular}{@{}rlrrr|rrr@{}}
    \toprule
          &       & \multicolumn{3}{c}{Individual} & \multicolumn{3}{c}{Aggregated} \\
\cmidrule{3-6}  \cmidrule{7-8}          & Devices & \multicolumn{1}{l}{Prec.} & \multicolumn{1}{l}{Recall} & \multicolumn{1}{l|}{F1 Sc.} & \multicolumn{1}{l}{Prec.} & \multicolumn{1}{l}{Recall} & \multicolumn{1}{l}{F1 Sc.} \\
    \hline
    \multicolumn{1}{c}{\multirow{31}[2]{*}{\begin{sideways}IP Devices\end{sideways}}} & Amcrest WiFi-Cam. & 0.999 & 0.967 & \cellcolor[rgb]{ .984,  .973,  .984}0.983 & 0.999 & 1.000 & \cellcolor[rgb]{ .984,  .984,  .996}1.000 \\
          & Amazon AE Dot & 0.913 & 0.972 & \cellcolor[rgb]{ .984,  .941,  .953}0.941 & 0.974 & 1.000 & \cellcolor[rgb]{ .984,  .976,  .988}0.987 \\
          & Amazon AE Spot & 0.420 & 0.852 & \cellcolor[rgb]{ .976,  .655,  .663}0.562 & 0.482 & 1.000 & \cellcolor[rgb]{ .98,  .745,  .753}0.650 \\
          & Amazon AE Studio & 0.794 & 0.867 & \cellcolor[rgb]{ .984,  .855,  .867}0.829 & 0.940 & 1.000 & \cellcolor[rgb]{ .984,  .965,  .976}0.969 \\
          & Amazon Plug & 0.997 & 0.999 & \cellcolor[rgb]{ .984,  .984,  .996}0.998 & 1.000 & 0.999 & \cellcolor[rgb]{ .984,  .984,  .996}1.000 \\
          & Arlo Base Station & 0.998 & 0.970 & \cellcolor[rgb]{ .984,  .976,  .988}0.984 & 1.000 & 1.000 & \cellcolor[rgb]{ .984,  .984,  .996}1.000 \\
          & Arlo Q Camera & 0.986 & 0.994 & \cellcolor[rgb]{ .984,  .98,  .992}0.990 & 1.000 & 1.000 & \cellcolor[rgb]{ .984,  .984,  .996}1.000 \\
          & Atomi Coff-Maker & 0.923 & 0.777 & \cellcolor[rgb]{ .984,  .867,  .878}0.844 & 1.000 & 0.998 & \cellcolor[rgb]{ .984,  .984,  .996}0.999 \\
          & Borun Camera & 0.986 & 0.974 & \cellcolor[rgb]{ .984,  .973,  .984}0.980 & 0.995 & 1.000 & \cellcolor[rgb]{ .984,  .984,  .996}0.997 \\
          & D-Link Mini Cam. & 0.988 & 0.980 & \cellcolor[rgb]{ .984,  .976,  .984}0.984 & 0.993 & 1.000 & \cellcolor[rgb]{ .984,  .984,  .996}0.996 \\
          & D-Link Water Sen. & 1.000 & 0.912 & \cellcolor[rgb]{ .984,  .953,  .965}0.954 & 1.000 & 1.000 & \cellcolor[rgb]{ .988,  .988,  1}1.000 \\
          & Eufy HomeBase 2 & 0.727 & 0.928 & \cellcolor[rgb]{ .984,  .847,  .855}0.815 & 0.995 & 1.000 & \cellcolor[rgb]{ .984,  .984,  .996}0.998 \\
          & Globe Lamp  & 0.707 & 0.609 & \cellcolor[rgb]{ .98,  .725,  .733}0.654 & 1.000 & 0.808 & \cellcolor[rgb]{ .984,  .914,  .925}0.894 \\
          & Google Nest Mini & 0.955 & 0.991 & \cellcolor[rgb]{ .984,  .965,  .976}0.973 & 0.983 & 1.000 & \cellcolor[rgb]{ .984,  .98,  .992}0.991 \\
          & Gosund Plug & 0.775 & 0.913 & \cellcolor[rgb]{ .984,  .863,  .875}0.839 & 0.938 & 1.000 & \cellcolor[rgb]{ .984,  .965,  .976}0.968 \\
          & Gosund Socket & 0.961 & 0.790 & \cellcolor[rgb]{ .984,  .886,  .898}0.867 & 1.000 & 0.986 & \cellcolor[rgb]{ .984,  .98,  .992}0.993 \\
          & HeimVision S Cam. & 1.000 & 0.971 & \cellcolor[rgb]{ .984,  .976,  .988}0.985 & 1.000 & 1.000 & \cellcolor[rgb]{ .984,  .984,  .996}1.000 \\
          & HeimVision Lamp & 0.861 & 0.609 & \cellcolor[rgb]{ .98,  .769,  .78}0.713 & 1.000 & 0.909 & \cellcolor[rgb]{ .984,  .953,  .965}0.952 \\
          & Home Eye Camera & 0.973 & 0.892 & \cellcolor[rgb]{ .984,  .933,  .945}0.931 & 1.000 & 0.999 & \cellcolor[rgb]{ .984,  .984,  .996}1.000 \\
          & Luohe Cam Dog & 0.723 & 0.804 & \cellcolor[rgb]{ .98,  .804,  .816}0.762 & 1.000 & 0.991 & \cellcolor[rgb]{ .984,  .984,  .996}0.996 \\
          & Nest Indoor Cam. & 0.999 & 0.999 & \cellcolor[rgb]{ .988,  .988,  1}0.999 & 1.000 & 0.999 & \cellcolor[rgb]{ .984,  .984,  .996}0.999 \\
          & Netatmo Camera & 0.950 & 0.991 & \cellcolor[rgb]{ .984,  .965,  .976}0.970 & 1.000 & 1.000 & \cellcolor[rgb]{ .984,  .984,  .996}1.000 \\
          & Netatmo Weather & 0.887 & 0.927 & \cellcolor[rgb]{ .984,  .914,  .925}0.906 & 0.999 & 1.000 & \cellcolor[rgb]{ .984,  .984,  .996}0.999 \\
          & Philips Hue Bridge & 0.997 & 0.992 & \cellcolor[rgb]{ .984,  .984,  .996}0.994 & 1.000 & 1.000 & \cellcolor[rgb]{ .984,  .984,  .996}1.000 \\
          & Ring Base Station & 0.656 & 0.149 & \cellcolor[rgb]{ .973,  .412,  .42}0.242 & 0.991 & 0.092 & \cellcolor[rgb]{ .973,  .412,  .42}0.167 \\
          & SIMCAM 1S & 0.996 & 0.995 & \cellcolor[rgb]{ .984,  .984,  .996}0.995 & 1.000 & 1.000 & \cellcolor[rgb]{ .984,  .984,  .996}1.000 \\
          & Smart Board & 0.859 & 0.239 & \cellcolor[rgb]{ .973,  .51,  .518}0.374 & 0.854 & 0.239 & \cellcolor[rgb]{ .973,  .553,  .561}0.373 \\
          & Sonos One Speaker & 0.583 & 0.916 & \cellcolor[rgb]{ .98,  .769,  .776}0.713 & 0.701 & 1.000 & \cellcolor[rgb]{ .984,  .863,  .875}0.824 \\
          & Teckin Plug & 0.698 & 0.652 & \cellcolor[rgb]{ .98,  .737,  .749}0.675 & 0.942 & 0.910 & \cellcolor[rgb]{ .984,  .933,  .945}0.926 \\
          & Yutron Plug & 0.669 & 0.889 & \cellcolor[rgb]{ .98,  .808,  .816}0.763 & 0.916 & 1.000 & \cellcolor[rgb]{ .984,  .957,  .969}0.956 \\
          & iRobot Roomba & 0.951 & 0.977 & \cellcolor[rgb]{ .984,  .961,  .973}0.964 & 0.976 & 1.000 & \cellcolor[rgb]{ .984,  .976,  .988}0.988 \\
    \hline
       \multicolumn{1}{r}{\multirow{9}[2]{*}{\begin{sideways}non-IP | Zigbee Devices\end{sideways}}} & AeoTec Button & 0.000 & 0.000 & \cellcolor[rgb]{ .973,  .412,  .42}0.000 & 0.000 & 0.000 & \cellcolor[rgb]{ .973,  .412,  .42}0.000 \\
          & AeoTec Mo Sen. & 0.056 & 0.240 & \cellcolor[rgb]{ .973,  .471,  .478}0.091 & 0.058 & 0.240 & \cellcolor[rgb]{ .973,  .471,  .478}0.093 \\
          & AeoTec MP Sen. & 0.506 & 0.514 & \cellcolor[rgb]{ .98,  .741,  .753}0.510 & 0.510 & 0.514 & \cellcolor[rgb]{ .98,  .741,  .753}0.512 \\
          & AeoTec WL Sen. & 0.020 & 0.125 & \cellcolor[rgb]{ .973,  .431,  .439}0.034 & 0.020 & 0.125 & \cellcolor[rgb]{ .973,  .431,  .439}0.034 \\
          & Philips Hue White & 0.839 & 0.942 & \cellcolor[rgb]{ .988,  .988,  1}0.888 & 0.840 & 0.942 & \cellcolor[rgb]{ .988,  .988,  1}0.888 \\
          & Sengled Smart Plug & 0.307 & 0.099 & \cellcolor[rgb]{ .973,  .506,  .514}0.149 & 0.308 & 0.099 & \cellcolor[rgb]{ .973,  .506,  .514}0.149 \\
          & SmartThings Button & 0.056 & 0.250 & \cellcolor[rgb]{ .973,  .471,  .478}0.092 & 0.057 & 0.250 & \cellcolor[rgb]{ .973,  .471,  .478}0.092 \\
          & SmartThings S. Bulb & 0.707 & 0.694 & \cellcolor[rgb]{ .984,  .863,  .875}0.700 & 0.766 & 0.694 & \cellcolor[rgb]{ .984,  .882,  .894}0.728 \\
          & Sonoff Smart Plug & 1.000 & 0.239 & \cellcolor[rgb]{ .976,  .659,  .671}0.386 & 1.000 & 0.239 & \cellcolor[rgb]{ .976,  .659,  .671}0.386 \\
    \hline
         
          & Macro avg & 0.761 & 0.740 & 0.726 & 0.831 & 0.801 & 0.788 \\
          & Weighted avg & 0.909 & 0.890 & 0.885 & 0.962 & 0.949 & 0.941 \\
           & Accuracy & \multicolumn{3}{c|}{0.890} & \multicolumn{3}{c}{0.949} \\
    \bottomrule
    \end{tabular}%
  \label{tab:non-ip-results}%
\end{table}%

Table~\ref{tab:non-ip-results} reveals that non-IP devices have minimal impact on the results of IP devices, showing almost identical performance to the previous evaluation. The only exception is the Sonos One Speaker, which exhibits a noticeable drop in F1 score, unrelated to the non-IP devices. This can be attributed to the above-mentioned anomaly in the Smart Board device.

On the other hand, the aggregation algorithm significantly improves the results for IP devices, while it has no effect on non-IP devices. This is due to the shared MAC address among Zigbee devices, which causes the aggregation algorithm to add them to the exception list and bypass aggregation for these devices.


Most of the non-IP devices show relatively low F1 scores. Nevertheless, the primary reason for these low results is likely the insufficient amount of data available for these devices. This performance issue was not encountered in the Aalto dataset, which contains more non-IP devices data, with a similar amount of data to the IP devices7. Table~\ref{tab:non-ip-packets} indicates a negative correlation between the number of packets and  F1 scores, further supporting the impact of insufficient data on performance. Additionally, it should be noted that the training and test data represent distinct states rather than being collected under normal conditions.


\section{Conclusions}\label{Conclusion}

This study validates the feature set and aggregation algorithm of the IoTDevID method using a new dataset. The dataset offers several advantages, including the presence of non-IP devices, multiple instances of the same device, normal usage data, and diverse usage profiles. The results demonstrate that models trained with \state{active} usage data outperform those trained with \state{idle} usage data, emphasizing the importance of data diversity in achieving better model performance.

The study showed successful external validation. It achieved impressive results with IP-only devices, achieving an F1 score of 92.50 for 31 device classes when evaluated in the more realistic scenario of devices undergoing \state{active} use. While non-IP devices faced challenges due to limited data availability, significant success was observed for devices with available data, showcasing the potential of the aggregation algorithm in accurately detecting non-IP devices.

In the original IoTDevID study, the UNSW dataset (32 IoT devices) achieved an F1 score of 93.70\%, similar to the results obtained here. The CIC-IoT-22 and UNSW datasets have similar usage data and device counts, with some structural differences. The Aalto dataset, with a lower number of devices, achieved a lower F1 score of 86\%. We attributed this to the dataset's abundance of devices performing similar tasks. Furthermore, the lack of usage data in the Aalto dataset may have contributed to the lower performance, as observed in our experiments on data quality.

This study highlights the importance of validation studies in assessing the robustness and generalizability of machine learning methods. The findings further contribute to the field of IoT device identification and provide insights into the impact of data diversity on model success.

Future research should focus on addressing the limitations related to insufficient data for non-IP devices, as well as exploring methods to enhance model performance in various scenarios. Additionally, investigating the scalability of the IoTDevID method to larger datasets and evaluating its applicability in real-world IoT environments would be valuable for practical implementation. Overall, this study serves as a foundation for further advancements in IoT device identification and security.

 \bibliographystyle{splncs04}
 \bibliography{references}
\newpage

\appendix \label{Appendices}
\begin{figure}[t]
	\centerline{\includegraphics[width=1\columnwidth]{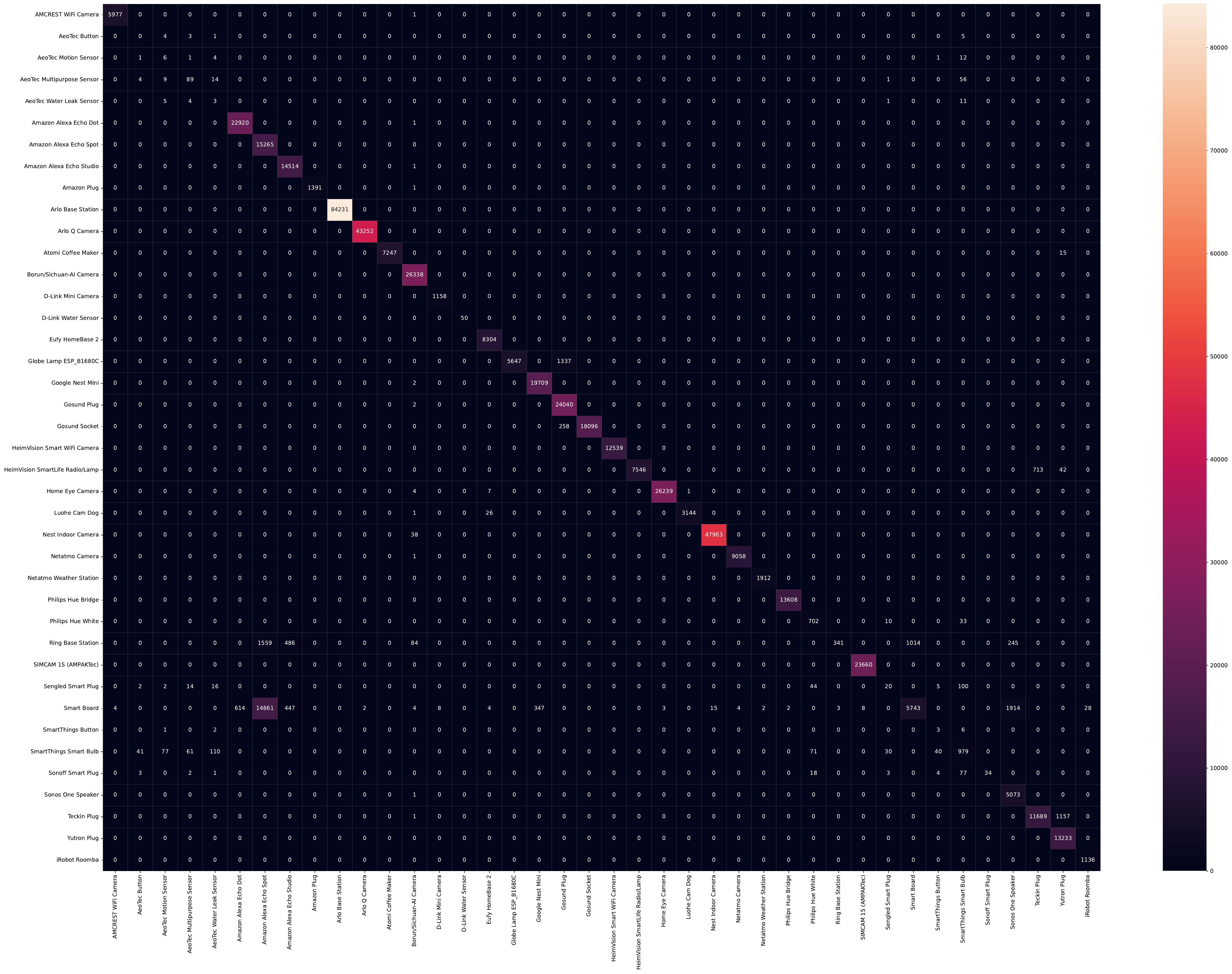}}
	\caption{Confusion matrix for case AA with non-IP devices.}
	\label{fig:cmson}
\end{figure}


\begin{landscape}
	\begin{table}[htbp]
		\centering
		\caption{Total number of packets generated by devices, and number of devices, in each session. Values scaled to 1000.}

  \resizebox{1.\textwidth}{!}{    \begin{tabular}{lccccccccccccccccccccccccccccccccccccccccccc}
  \hline
    \begin{turn}{-90}ML  Input Data  Type\end{turn} & \begin{turn}{-90}Date of Data Collection\end{turn} & \begin{turn}{-90}State Type (\state{active}/\state{idle})\end{turn} & \begin{turn}{-90}AMCREST WiFi Camera\end{turn} & \begin{turn}{-90}Amazon Alexa Echo Dot 1\end{turn} & \begin{turn}{-90}Amazon Alexa Echo Dot 2\end{turn} & \begin{turn}{-90}Amazon Alexa Echo Spot\end{turn} & \begin{turn}{-90}Amazon Alexa Echo Studio\end{turn} & \begin{turn}{-90}Amazon Plug\end{turn} & \begin{turn}{-90}Arlo Base Station\end{turn} & \begin{turn}{-90}Arlo Q Camera\end{turn} & \begin{turn}{-90}Atomi Coffee Maker\end{turn} & \begin{turn}{-90}Borun/Sichuan-AI Camera\end{turn} & \begin{turn}{-90}D-Link Water Sensor\end{turn} & \begin{turn}{-90}D-Link Mini Camera\end{turn} & \begin{turn}{-90}Eufy HomeBase 2\end{turn} & \begin{turn}{-90}Globe Lamp ESP\_B1680C\end{turn} & \begin{turn}{-90}Google Nest Mini\end{turn} & \begin{turn}{-90}Gosund ESP\_032979 Plug\end{turn} & \begin{turn}{-90}Gosund ESP\_039AAF Sckt\end{turn} & \begin{turn}{-90}Gosund ESP\_0C3994 Plug\end{turn} & \begin{turn}{-90}Gosund ESP\_10098F Sckt\end{turn} & \begin{turn}{-90}Gosund ESP\_10ACD8 Plug\end{turn} & \begin{turn}{-90}Gosund ESP\_147FF9 Plug\end{turn} & \begin{turn}{-90}Gosund ESP\_1ACEE1  Sckt\end{turn} & \begin{turn}{-90}HeimVision Smrt WiFiCam\end{turn} & \begin{turn}{-90}HeimVision SL R-Lamp\end{turn} & \begin{turn}{-90}Home Eye Camera\end{turn} & \begin{turn}{-90}LG Smart TV\end{turn} & \begin{turn}{-90}Luohe Cam Dog\end{turn} & \begin{turn}{-90}Nest Indoor Camera\end{turn} & \begin{turn}{-90}Netatmo Camera\end{turn} & \begin{turn}{-90}Netatmo Weather Station\end{turn} & \begin{turn}{-90}Philips Hue Bridge\end{turn} & \begin{turn}{-90}Ring Base Station\end{turn} & \begin{turn}{-90}SIMCAM 1S (AMPAKTec)\end{turn} & \begin{turn}{-90}Smart Board\end{turn} & \begin{turn}{-90}Sonos One Speaker\end{turn} & \begin{turn}{-90}Teckin Plug 1\end{turn} & \begin{turn}{-90}Teckin Plug 2\end{turn} & \begin{turn}{-90}Yutron Plug 1\end{turn} & \begin{turn}{-90}Yutron Plug 2\end{turn} & \begin{turn}{-90}iRobot Roomba\end{turn} & \begin{turn}{-90}\textbf{Total Packet Numbers}\end{turn} \\\hline
    \rowcolor[rgb]{ .851,  .851,  .851} Train & 20211102 & \state{active} & 6.68  & 6.91  & 6.84  & 14.94 & 13.20 & 1.47  & 12.63 & 21.08 & 6.94  & 24.04 & 0.00  & 1.08  & 8.08  & 6.75  & 20.59 & 6.65  & 6.62  & 6.63  & 0.00  & 6.66  & 6.63  & 6.64  & 14.30 & 7.88  & 22.48 & 0     & 3.03  & 33.40 & 9.59  & 1.81  & 12.50 & 3.37  & 21.45 & 0.40  & 5.18  & 6.68  & 6.34  & 6.35  & 6.68  & 1.12  & \cellcolor[rgb]{ .98,  .6,  .455}353.62 \\
    Train & 20211103 & \state{active} & 5.54  & 7.10  & 6.98  & 15.19 & 13.16 & 1.48  & 4.93  & 18.81 & 6.93  & 24.42 & 0.00  & 1.10  & 8.07  & 6.75  & 24.46 & 6.75  & 6.74  & 6.75  & 0.00  & 6.75  & 6.74  & 6.75  & 14.29 & 7.88  & 25.28 & 600   & 2.92  & 24.23 & 12.21 & 1.87  & 12.67 & 3.26  & 21.08 & 0.55  & 8.71  & 6.62  & 6.29  & 6.53  & 6.65  & 1.21  & \cellcolor[rgb]{ .855,  .882,  .51}947.57 \\
    \rowcolor[rgb]{ .851,  .851,  .851} Train & 20211105 & \state{active} & 5.52  & 7.00  & 6.93  & 13.72 & 13.12 & 1.48  & 4.95  & 21.54 & 6.94  & 26.20 & 0.00  & 1.11  & 8.07  & 6.74  & 21.27 & 6.74  & 6.74  & 6.75  & 0.00  & 6.75  & 6.75  & 6.76  & 13.82 & 7.88  & 29.66 & 0     & 2.78  & 23.78 & 26.15 & 1.70  & 14.03 & 3.21  & 20.91 & 0.38  & 4.97  & 6.61  & 6.46  & 6.63  & 6.06  & 1.05  & \cellcolor[rgb]{ .984,  .631,  .459}361.14 \\
    Train & 20211108 & \state{active} & 5.63  & 10.73 & 10.92 & 14.99 & 13.06 & 1.47  & 45.08 & 78.96 & 6.94  & 26.05 & 0.00  & 1.11  & 8.08  & 6.74  & 21.11 & 6.70  & 6.33  & 6.36  & 3.29  & 6.37  & 6.34  & 6.35  & 13.78 & 7.90  & 37.03 & 0     & 2.98  & 84.38 & 10.19 & 1.82  & 13.81 & 3.47  & 21.96 & 0.43  & 4.97  & 6.59  & 6.07  & 6.61  & 6.12  & 1.06  & \cellcolor[rgb]{ .976,  .918,  .518}521.74 \\
    \rowcolor[rgb]{ .851,  .851,  .851} Train & 20211109 & \state{active} & 6.09  & 11.44 & 10.94 & 14.72 & 13.17 & 1.47  & 45.97 & 47.26 & 6.93  & 24.37 & 0.00  & 1.09  & 7.97  & 6.75  & 21.08 & 6.75  & 6.74  & 6.75  & 6.75  & 6.75  & 6.75  & 6.76  & 13.76 & 7.88  & 25.97 & 0     & 3.03  & 43.68 & 9.17  & 1.77  & 12.47 & 3.34  & 21.65 & 0.44  & 4.99  & 6.62  & 6.08  & 6.64  & 6.07  & 1.05  & \cellcolor[rgb]{ 1,  .922,  .518}441.11 \\
    Train & 20211110 & \state{active} & 6.74  & 44.07 & 48.50 & 19.88 & 14.95 & 2.39  & 22.25 & 37.38 & 6.93  & 24.20 & 0.00  & 1.09  & 7.98  & 6.74  & 17.82 & 6.73  & 6.73  & 6.74  & 6.75  & 6.75  & 0.00  & 6.75  & 13.79 & 7.86  & 33.05 & 0     & 3.01  & 26.28 & 9.00  & 1.82  & 13.79 & 3.27  & 21.15 & 6.15  & 6.87  & 6.09  & 6.11  & 6.46  & 6.49  & 1.16  & \cellcolor[rgb]{ .988,  .922,  .518}473.70 \\
    \rowcolor[rgb]{ .851,  .851,  .851} Train & 20211112 & \state{active} & 5.65  & 14.08 & 40.82 & 15.70 & 19.46 & 1.47  & 4.89  & 37.48 & 6.92  & 24.09 & 0.00  & 1.09  & 8.06  & 6.73  & 25.21 & 6.73  & 6.73  & 6.74  & 6.74  & 6.75  & 0.00  & 6.73  & 13.74 & 7.86  & 39.65 & 0     & 3.02  & 29.18 & 10.53 & 1.91  & 15.19 & 3.28  & 21.48 & 8.36  & 6.98  & 6.25  & 6.26  & 6.56  & 6.39  & 1.19  & \cellcolor[rgb]{ 1,  .922,  .518}439.88 \\
    Train & 20211115 & \state{active} & 5.66  & 12.14 & 11.83 & 14.93 & 14.21 & 1.82  & 110   & 55.18 & 7.14  & 17.26 & 0.00  & 1.17  & 8.15  & 6.88  & 24.21 & 6.87  & 6.89  & 6.89  & 6.90  & 6.89  & 6.88  & 6.90  & 13.61 & 8.07  & 23.00 & 0     & 3.01  & 67.01 & 9.43  & 1.83  & 17.50 & 3.85  & 21.47 & 3.04  & 4.25  & 6.36  & 6.28  & 6.58  & 6.59  & 1.21  & \cellcolor[rgb]{ .969,  .914,  .518}541.98 \\
    \rowcolor[rgb]{ .851,  .851,  .851} Train & 20211116 & \state{active} & 5.43  & 11.33 & 10.96 & 13.68 & 13.82 & 1.47  & 11.09 & 98.64 & 6.93  & 22.57 & 0.00  & 1.11  & 8.01  & 6.74  & 21.04 & 6.73  & 6.75  & 6.75  & 6.77  & 6.74  & 6.73  & 6.75  & 13.93 & 7.91  & 27.98 & 0     & 3.02  & 84.55 & 14.13 & 1.90  & 15.72 & 3.32  & 22.03 & 8.06  & 3.84  & 6.11  & 6.01  & 6.26  & 6.25  & 1.17  & \cellcolor[rgb]{ .98,  .918,  .518}512.23 \\
    Train & 20211117 & \state{active} & 5.36  & 23.47 & 22.58 & 15.33 & 58.61 & 1.47  & 4.88  & 44.92 & 6.93  & 24.42 & 0.00  & 1.12  & 7.95  & 6.73  & 14.10 & 6.73  & 6.73  & 6.75  & 6.75  & 6.74  & 6.73  & 6.73  & 13.66 & 7.88  & 27.78 & 0     & 3.02  & 29.93 & 9.54  & 1.91  & 13.70 & 3.25  & 21.23 & 3.99  & 4.84  & 6.15  & 6.01  & 6.14  & 6.35  & 1.10  & \cellcolor[rgb]{ .996,  .922,  .518}451.52 \\
    \rowcolor[rgb]{ .851,  .851,  .851} Test  & 20211118 & \state{active} & 5.97  & 11.78 & 11.35 & 13.33 & 14.21 & 1.47  & 4.81  & 4.97  & 6.94  & 24.18 & 0.00  & 1.11  & 8.12  & 6.73  & 20.55 & 6.74  & 6.74  & 6.75  & 6.75  & 6.74  & 6.74  & 6.74  & 13.66 & 7.88  & 21.26 & 0     & 3.02  & 29.55 & 4.31  & 1.88  & 12.86 & 3.20  & 21.01 & 3.63  & 3.89  & 6.26  & 6.21  & 6.28  & 6.38  & 1.34  & \cellcolor[rgb]{ .976,  .525,  .439}335.33 \\
    Test  & 20211119 & \state{active} & 5.54  & 11.83 & 11.15 & 13.41 & 14.22 & 1.47  & 4.91  & 3.64  & 6.92  & 24.12 & 0.00  & 1.11  & 8.07  & 6.73  & 20.25 & 6.73  & 6.73  & 6.75  & 6.76  & 6.74  & 6.73  & 6.73  & 0.00  & 7.87  & 21.08 & 0     & 3.02  & 29.57 & 4.10  & 1.85  & 12.70 & 3.21  & 20.92 & 1.37  & 3.82  & 6.14  & 6.00  & 6.13  & 6.33  & 1.12  & \cellcolor[rgb]{ .973,  .439,  .424}315.75 \\
    \rowcolor[rgb]{ .851,  .851,  .851} Test  & 20211122 & \state{active} & 5.62  & 9.54  & 9.35  & 14.91 & 13.49 & 1.47  & 15.10 & 58.72 & 6.94  & 24.12 & 0.00  & 1.12  & 8.15  & 6.73  & 20.72 & 6.71  & 6.71  & 6.72  & 6.74  & 6.72  & 6.72  & 6.71  & 13.64 & 7.87  & 22.20 & 0     & 3.02  & 78.49 & 9.49  & 1.82  & 13.54 & 3.32  & 21.26 & 3.82  & 3.87  & 6.13  & 6.05  & 6.39  & 6.38  & 1.12  & \cellcolor[rgb]{ .996,  .922,  .518}451.39 \\
    Test  & 20211123 & \state{active} & 6.28  & 11.53 & 10.46 & 15.34 & 13.58 & 1.41  & 7.82  & 58.50 & 7.10  & 24.76 & 0.00  & 1.15  & 8.12  & 6.89  & 20.72 & 6.88  & 6.87  & 6.90  & 6.91  & 6.89  & 6.88  & 6.88  & 13.90 & 8.07  & 26.72 & 0     & 3.09  & 74.48 & 11.61 & 1.85  & 12.97 & 3.33  & 28.15 & 5.24  & 3.93  & 6.25  & 6.18  & 6.57  & 6.59  & 1.06  & \cellcolor[rgb]{ .992,  .922,  .518}461.84 \\
    \rowcolor[rgb]{ .851,  .851,  .851} Test  & 20211124 & \state{active} & 5.74  & 11.22 & 10.28 & 14.61 & 13.34 & 1.47  & 4.90  & 17.54 & 6.93  & 24.88 & 0.08  & 1.10  & 7.96  & 6.73  & 20.23 & 6.73  & 6.73  & 6.74  & 6.75  & 6.74  & 6.73  & 6.73  & 0.09  & 7.88  & 21.73 & 0     & 3.01  & 23.02 & 6.47  & 1.86  & 12.57 & 3.29  & 27.77 & 4.74  & 4.11  & 6.10  & 6.23  & 6.46  & 6.34  & 1.03  & \cellcolor[rgb]{ .976,  .529,  .439}336.83 \\
    Test  & 20211125 & \state{active} & 5.76  & 11.37 & 10.25 & 16.00 & 14.08 & 1.55  & 5.15  & 34.63 & 6.94  & 24.23 & 0.00  & 1.09  & 8.07  & 6.73  & 20.68 & 6.73  & 6.73  & 6.76  & 6.76  & 6.76  & 6.74  & 6.73  & 29.56 & 7.88  & 21.17 & 0     & 3.03  & 22.86 & 5.75  & 1.90  & 14.36 & 3.30  & 21.16 & 6.04  & 5.75  & 6.10  & 6.18  & 6.30  & 6.31  & 1.18  & \cellcolor[rgb]{ .988,  .725,  .478}382.52 \\
    \rowcolor[rgb]{ .851,  .851,  .851} Test  & 20211126 & \state{active} & 7.74  & 16.24 & 14.83 & 20.22 & 17.21 & 0.52  & 326   & 148.99 & 9.20  & 32.01 & 0.00  & 1.65  & 11.30 & 8.93  & 26.73 & 8.93  & 8.93  & 8.94  & 8.92  & 8.95  & 8.97  & 8.94  & 18.23 & 10.45 & 42.15 & 30    & 4.05  & 39.26 & 19.26 & 2.47  & 17.65 & 5.69  & 29.41 & 212   & 6.21  & 8.18  & 8.72  & 8.65  & 8.37  & 1.56  & \cellcolor[rgb]{ .792,  .863,  .506}1176.98 \\
    Test  & 20211206 & \state{active} & 6.50  & 11.06 & 10.68 & 15.59 & 15.76 & 1.56  & 24.18 & 33.81 & 6.94  & 32.40 & 0.00  & 1.09  & 8.79  & 6.73  & 18.50 & 6.75  & 6.73  & 6.74  & 6.74  & 6.78  & 6.73  & 6.74  & 13.81 & 7.88  & 18.93 & 0     & 2.97  & 69.60 & 11.83 & 1.77  & 12.83 & 3.41  & 21.78 & 0.77  & 4.66  & 6.08  & 6.32  & 6.55  & 6.55  & 1.20  & \cellcolor[rgb]{ 1,  .922,  .518}437.73 \\
    \rowcolor[rgb]{ .851,  .851,  .851} Test  & 20211207 & \state{active} & 5.66  & 10.97 & 10.55 & 15.57 & 14.44 & 1.52  & 438   & 41.61 & 6.93  & 27.73 & 0.00  & 1.09  & 6.53  & 6.73  & 14.97 & 2.72  & 6.27  & 2.67  & 2.69  & 2.67  & 2.66  & 2.67  & 15.55 & 7.89  & 29.16 & 1805  & 3.02  & 27.10 & 9.89  & 1.82  & 14.18 & 3.81  & 21.62 & 0.43  & 11.62 & 6.73  & 6.25  & 6.68  & 6.78  & 1.05  & \cellcolor[rgb]{ .388,  .745,  .482}2603.11 \\
    Test  & 20211208 & \state{active} & 6.16  & 10.66 & 10.59 & 15.34 & 13.63 & 1.52  & 13.70 & 29.37 & 6.94  & 25.18 & 0.00  & 1.12  & 8.03  & 6.74  & 14.31 & 1.04  & 0.98  & 0.98  & 1.05  & 0.98  & 0.97  & 0.98  & 5.44  & 7.91  & 39.17 & 0     & 3.02  & 89.40 & 9.71  & 1.79  & 12.85 & 3.46  & 21.97 & 0.51  & 3.89  & 6.64  & 6.07  & 6.59  & 6.54  & 1.06  & \cellcolor[rgb]{ .992,  .78,  .49}396.29 \\
    \rowcolor[rgb]{ .851,  .851,  .851} Train & 20211223 & \state{active} & 6.26  & 11.73 & 11.49 & 13.33 & 13.33 & 1.46  & 4.93  & 3.79  & 6.92  & 24.79 & 0.00  & 1.08  & 7.99  & 6.72  & 19.71 & 6.75  & 6.72  & 6.72  & 6.72  & 6.72  & 6.72  & 6.72  & 13.57 & 7.90  & 17.00 & 0     & 3.02  & 23.24 & 3.78  & 1.77  & 12.81 & 3.17  & 20.89 & 0.74  & 4.98  & 6.81  & 6.67  & 6.74  & 6.46  & 1.20  & \cellcolor[rgb]{ .973,  .463,  .427}321.38 \\
    Train & 20211225 & \state{active} & 5.68  & 16.13 & 11.76 & 14.88 & 14.39 & 1.46  & 13.16 & 23.09 & 6.93  & 25.21 & 0.00  & 1.08  & 8.01  & 6.73  & 19.97 & 6.74  & 6.72  & 6.72  & 6.72  & 6.72  & 6.73  & 6.72  & 13.52 & 7.89  & 19.59 & 0     & 3.02  & 16.77 & 7.10  & 1.88  & 12.95 & 3.20  & 21.15 & 2.71  & 11.82 & 6.80  & 6.68  & 6.75  & 6.35  & 1.04  & \cellcolor[rgb]{ .984,  .647,  .463}364.74 \\
    \rowcolor[rgb]{ .851,  .851,  .851} Train & 20211228 & \state{active} & 5.45  & 11.56 & 11.93 & 20.39 & 15.07 & 1.46  & 8.51  & 23.50 & 6.92  & 24.60 & 0.00  & 1.07  & 8.45  & 6.72  & 13.55 & 6.75  & 6.72  & 6.72  & 6.72  & 6.72  & 6.72  & 6.72  & 13.52 & 7.89  & 19.59 & 0     & 3.02  & 24.72 & 5.67  & 1.81  & 7.08  & 3.23  & 21.19 & 0.61  & 44.11 & 6.79  & 6.68  & 6.64  & 6.17  & 1.10  & \cellcolor[rgb]{ .988,  .737,  .482}386.08 \\
    Train & 20220103 & \state{active} & 5.59  & 11.72 & 9.36  & 14.45 & 13.35 & 1.50  & 4.91  & 2.75  & 6.93  & 25.43 & 0.00  & 1.08  & 8.07  & 6.72  & 14.93 & 6.73  & 6.72  & 6.74  & 6.72  & 6.73  & 6.75  & 6.73  & 14.93 & 7.86  & 16.75 & 0     & 0.00  & 25.41 & 3.74  & 1.68  & 7.07  & 3.16  & 20.91 & 0.48  & 4.92  & 6.68  & 6.37  & 6.61  & 6.60  & 1.39  & \cellcolor[rgb]{ .973,  .412,  .42}308.43 \\
    \rowcolor[rgb]{ .851,  .851,  .851} Test  & 20211102 & \state{idle}  & 7.92  & 8.94  & 8.90  & 19.26 & 121.47 & 2.00  & 6.85  & 3.82  & 9.54  & 32.99 & 0.00  & 1.46  & 11.04 & 9.26  & 29.43 & 9.26  & 9.25  & 9.26  & 0.00  & 9.27  & 9.25  & 9.26  & 20.23 & 10.84 & 28.76 & 0     & 4.15  & 6.63  & 7.35  & 2.47  & 17.05 & 4.79  & 28.62 & 0.35  & 6.89  & 9.08  & 8.70  & 8.82  & 9.08  & 1.43  & \cellcolor[rgb]{ .98,  .918,  .518}503.68 \\
    Test  & 20211103 & \state{idle}  & 7.85  & 9.09  & 9.11  & 18.78 & 19.42 & 2.05  & 6.83  & 3.82  & 9.52  & 33.06 & 0.00  & 1.46  & 11.12 & 9.25  & 30.55 & 9.26  & 9.27  & 9.28  & 0.00  & 9.27  & 9.26  & 9.27  & 20.05 & 10.83 & 28.81 & 0     & 3.66  & 6.48  & 7.27  & 2.46  & 16.93 & 4.70  & 28.64 & 0.20  & 6.82  & 9.06  & 8.71  & 8.95  & 9.07  & 1.42  & \cellcolor[rgb]{ .992,  .804,  .494}401.55 \\
    \rowcolor[rgb]{ .851,  .851,  .851} Test  & 20211104 & \state{idle}  & 8.60  & 9.32  & 9.36  & 18.90 & 19.41 & 2.09  & 6.89  & 3.94  & 9.51  & 34.96 & 0.00  & 1.47  & 11.23 & 9.26  & 22.58 & 9.27  & 9.26  & 9.26  & 0.00  & 9.28  & 9.26  & 9.28  & 19.78 & 10.83 & 28.76 & 0     & 3.80  & 7.69  & 7.69  & 2.53  & 16.88 & 4.73  & 28.60 & 0.24  & 6.90  & 8.91  & 8.84  & 8.88  & 8.56  & 1.51  & \cellcolor[rgb]{ .992,  .792,  .49}398.29 \\
    Test  & 20211108 & \state{idle}  & 7.97  & 15.31 & 15.73 & 17.53 & 19.50 & 2.05  & 6.81  & 3.81  & 9.55  & 35.11 & 0.00  & 1.52  & 10.97 & 9.27  & 29.83 & 9.26  & 9.25  & 9.27  & 9.27  & 9.26  & 9.26  & 9.29  & 19.07 & 10.85 & 28.78 & 0     & 4.15  & 20.06 & 5.12  & 2.51  & 16.89 & 4.69  & 28.64 & 0.05  & 6.82  & 9.11  & 8.36  & 9.11  & 8.38  & 1.52  & \cellcolor[rgb]{ 1,  .922,  .518}433.88 \\
    \rowcolor[rgb]{ .851,  .851,  .851} Test  & 20211110 & \state{idle}  & 8.43  & 15.27 & 15.38 & 19.42 & 17.88 & 1.98  & 6.84  & 51.55 & 9.54  & 34.98 & 0.00  & 1.48  & 11.27 & 9.25  & 31.72 & 9.25  & 9.25  & 9.27  & 9.27  & 9.27  & 0.00  & 9.25  & 18.83 & 10.82 & 29.00 & 0     & 4.15  & 8.06  & 5.17  & 2.51  & 17.11 & 4.68  & 28.63 & 3.87  & 6.86  & 8.39  & 8.30  & 9.12  & 8.93  & 1.43  & \cellcolor[rgb]{ .992,  .922,  .518}466.40 \\
    Test  & 20211112 & \state{idle}  & 8.53  & 15.11 & 14.88 & 17.22 & 17.35 & 2.00  & 6.80  & 47.37 & 9.54  & 34.99 & 0.00  & 1.47  & 11.01 & 9.25  & 40.40 & 9.25  & 9.25  & 9.27  & 9.27  & 9.28  & 9.28  & 9.25  & 18.82 & 10.82 & 28.81 & 0     & 4.15  & 11.71 & 5.12  & 2.55  & 24.88 & 4.72  & 28.64 & 3.30  & 6.93  & 8.41  & 8.40  & 9.03  & 8.70  & 1.43  & \cellcolor[rgb]{ .984,  .918,  .518}487.15 \\
    \rowcolor[rgb]{ .851,  .851,  .851} Test  & 20211115 & \state{idle}  & 7.44  & 16.37 & 15.87 & 17.82 & 18.01 & 1.99  & 10.25 & 3.95  & 9.55  & 22.25 & 0.00  & 1.48  & 11.08 & 9.26  & 28.80 & 9.26  & 9.25  & 9.29  & 9.30  & 9.29  & 9.28  & 9.25  & 18.84 & 10.82 & 28.50 & 0     & 4.12  & 41.22 & 5.11  & 2.58  & 24.87 & 4.65  & 28.63 & 1.41  & 5.31  & 8.48  & 8.38  & 8.72  & 8.83  & 1.43  & \cellcolor[rgb]{ .996,  .922,  .518}450.91 \\
    Test  & 20211116 & \state{idle}  & 7.50  & 16.55 & 15.58 & 17.92 & 18.09 & 1.99  & 6.81  & 3.84  & 9.54  & 34.99 & 0.00  & 1.50  & 11.12 & 9.25  & 32.30 & 9.25  & 9.25  & 9.28  & 9.27  & 9.27  & 9.26  & 9.25  & 18.83 & 10.81 & 28.44 & 0     & 4.15  & 40.95 & 5.20  & 2.47  & 17.21 & 4.69  & 28.64 & 0.00  & 5.33  & 8.43  & 8.29  & 8.56  & 8.71  & 1.50  & \cellcolor[rgb]{ .996,  .922,  .518}454.01 \\
    \rowcolor[rgb]{ .851,  .851,  .851} Test  & 20211117 & \state{idle}  & 7.56  & 16.46 & 16.16 & 17.94 & 18.52 & 2.01  & 6.86  & 3.84  & 9.54  & 33.14 & 0.00  & 1.51  & 11.02 & 9.25  & 20.19 & 9.25  & 9.25  & 9.27  & 9.27  & 9.27  & 9.27  & 9.25  & 18.87 & 10.83 & 28.82 & 0     & 4.16  & 40.84 & 5.26  & 2.59  & 17.20 & 4.75  & 28.65 & 0.00  & 21.86 & 8.41  & 8.24  & 8.68  & 8.69  & 1.51  & \cellcolor[rgb]{ .992,  .922,  .518}458.17 \\
    Train & 20211118 & \state{idle}  & 7.56  & 16.78 & 16.06 & 18.00 & 18.49 & 2.00  & 6.78  & 3.80  & 9.53  & 33.10 & 0.00  & 1.47  & 11.01 & 9.25  & 28.82 & 9.25  & 9.25  & 9.28  & 9.28  & 9.28  & 9.26  & 9.25  & 13.83 & 10.81 & 28.82 & 0     & 4.15  & 40.66 & 5.16  & 2.63  & 17.17 & 4.72  & 28.64 & 3.38  & 5.36  & 8.42  & 8.27  & 8.64  & 8.67  & 1.48  & \cellcolor[rgb]{ .996,  .922,  .518}448.29 \\
    \rowcolor[rgb]{ .851,  .851,  .851} Train & 20211119 & \state{idle}  & 8.25  & 16.65 & 16.41 & 18.14 & 18.41 & 2.00  & 6.90  & 5.31  & 9.54  & 34.98 & 0.00  & 1.46  & 11.37 & 9.25  & 28.66 & 9.25  & 9.25  & 9.26  & 9.29  & 9.28  & 9.28  & 9.25  & 4.64  & 10.82 & 29.03 & 0     & 4.15  & 40.78 & 5.21  & 2.50  & 17.28 & 4.71  & 28.64 & 1.29  & 5.42  & 8.42  & 8.26  & 8.56  & 8.67  & 1.42  & \cellcolor[rgb]{ 1,  .922,  .518}441.97 \\
    Train & 20211122 & \state{idle}  & 7.68  & 14.84 & 12.67 & 20.86 & 17.41 & 2.05  & 6.81  & 3.80  & 9.54  & 33.25 & 0.00  & 1.48  & 10.96 & 9.25  & 28.37 & 9.25  & 9.25  & 9.28  & 9.30  & 9.29  & 9.27  & 9.25  & 18.81 & 10.81 & 28.94 & 0     & 4.16  & 40.82 & 5.05  & 2.56  & 17.25 & 4.70  & 28.63 & 1.34  & 5.32  & 8.35  & 8.39  & 8.84  & 8.87  & 1.46  & \cellcolor[rgb]{ .996,  .922,  .518}448.13 \\
    \rowcolor[rgb]{ .851,  .851,  .851} Train & 20211124 & \state{idle}  & 7.75  & 15.58 & 14.15 & 18.74 & 17.16 & 2.01  & 6.83  & 4.34  & 9.53  & 33.27 & 0.07  & 1.49  & 11.17 & 9.24  & 28.26 & 9.25  & 9.25  & 9.29  & 9.28  & 9.28  & 9.27  & 9.25  & 18.79 & 10.82 & 29.05 & 0     & 4.15  & 7.00  & 5.14  & 2.48  & 17.18 & 4.65  & 37.53 & 4.55  & 5.38  & 8.44  & 8.38  & 8.72  & 8.63  & 1.43  & \cellcolor[rgb]{ .996,  .91,  .514}426.78 \\
    Train & 20211125 & \state{idle}  & 7.95  & 15.88 & 14.38 & 18.76 & 17.59 & 2.04  & 6.82  & 3.81  & 9.54  & 33.25 & 0.04  & 1.49  & 10.98 & 9.25  & 28.65 & 9.25  & 9.25  & 9.29  & 9.29  & 9.29  & 9.27  & 9.25  & 21.05 & 10.82 & 28.61 & 0     & 4.16  & 6.82  & 5.07  & 2.55  & 17.18 & 4.68  & 28.64 & 0.48  & 5.37  & 8.40  & 8.32  & 8.44  & 8.55  & 1.43  & \cellcolor[rgb]{ .996,  .867,  .506}415.90 \\
    \rowcolor[rgb]{ .851,  .851,  .851} Train & 20211126 & \state{idle}  & 8.95  & 16.37 & 15.58 & 21.80 & 19.96 & 0.00  & 6.85  & 4.38  & 9.55  & 34.99 & 0.07  & 1.50  & 11.00 & 9.25  & 28.43 & 9.25  & 9.26  & 9.28  & 9.29  & 9.29  & 9.27  & 9.25  & 18.91 & 10.82 & 29.32 & 0     & 4.18  & 21.80 & 5.00  & 2.50  & 17.21 & 4.69  & 28.62 & 4.77  & 5.38  & 8.37  & 8.60  & 8.63  & 8.62  & 1.42  & \cellcolor[rgb]{ 1,  .922,  .518}442.39 \\
    Train & 20211129 & \state{idle}  & 8.49  & 15.30 & 14.39 & 19.23 & 19.33 & 1.97  & 7.04  & 3.97  & 9.61  & 33.24 & 0.04  & 1.50  & 11.01 & 9.32  & 19.83 & 9.34  & 9.34  & 9.38  & 9.34  & 9.34  & 9.34  & 9.33  & 20.23 & 10.88 & 23.21 & 0     & 4.17  & 6.52  & 5.02  & 2.59  & 17.31 & 0.00  & 28.73 & 0.39  & 5.40  & 8.51  & 8.62  & 9.00  & 9.04  & 1.48  & \cellcolor[rgb]{ .992,  .8,  .494}400.76 \\
    \rowcolor[rgb]{ .851,  .851,  .851} Train & 20211130 & \state{idle}  & 7.79  & 15.77 & 14.29 & 19.18 & 17.51 & 2.00  & 10.01 & 3.80  & 9.53  & 33.67 & 0.07  & 1.49  & 11.16 & 9.25  & 19.26 & 0.00  & 9.25  & 9.25  & 9.26  & 9.25  & 9.25  & 9.25  & 18.75 & 10.84 & 23.20 & 0     & 4.15  & 6.65  & 5.08  & 2.46  & 17.27 & 4.70  & 28.62 & 0.38  & 5.34  & 8.30  & 8.86  & 8.25  & 8.55  & 1.42  & \cellcolor[rgb]{ .988,  .769,  .486}393.10 \\
    Train & 20211201 & \state{idle}  & 8.90  & 15.49 & 14.60 & 18.90 & 17.48 & 2.01  & 6.85  & 3.80  & 9.53  & 35.77 & 0.04  & 1.49  & 11.03 & 9.24  & 21.10 & 9.26  & 9.25  & 9.25  & 9.26  & 9.25  & 9.25  & 9.25  & 18.75 & 10.84 & 23.24 & 0     & 4.14  & 7.39  & 5.19  & 2.56  & 17.26 & 4.71  & 28.63 & 0.44  & 5.37  & 8.28  & 8.87  & 8.27  & 8.71  & 1.43  & \cellcolor[rgb]{ .992,  .82,  .498}405.04 \\
    \rowcolor[rgb]{ .851,  .851,  .851} Train & 20211202 & \state{idle}  & 7.77  & 14.71 & 14.15 & 19.08 & 17.49 & 1.97  & 6.90  & 3.80  & 9.53  & 35.70 & 0.11  & 2.53  & 11.16 & 9.25  & 20.06 & 9.28  & 9.25  & 9.25  & 9.27  & 9.24  & 9.25  & 9.25  & 18.84 & 10.85 & 23.69 & 0     & 4.13  & 7.83  & 5.15  & 2.52  & 17.25 & 4.68  & 28.60 & 0.50  & 5.34  & 8.26  & 8.80  & 9.07  & 8.50  & 1.43  & \cellcolor[rgb]{ .992,  .816,  .494}404.43 \\
    Train & 20211203 & \state{idle}  & 8.44  & 14.90 & 14.75 & 18.24 & 17.32 & 1.99  & 6.86  & 3.82  & 9.53  & 35.76 & 0.07  & 1.48  & 11.34 & 9.25  & 28.62 & 9.27  & 9.25  & 9.25  & 9.27  & 9.25  & 9.25  & 9.33  & 18.80 & 10.85 & 23.51 & 0     & 4.16  & 8.64  & 5.14  & 2.47  & 17.70 & 4.67  & 28.63 & 14.91 & 5.20  & 8.44  & 8.60  & 9.10  & 8.93  & 1.00  & \cellcolor[rgb]{ .996,  .918,  .514}427.99 \\
    \rowcolor[rgb]{ .851,  .851,  .851} Train & 20211207 & \state{idle}  & 8.05  & 15.23 & 14.13 & 18.63 & 19.22 & 2.01  & 6.91  & 3.81  & 9.53  & 35.77 & 0.04  & 1.48  & 11.28 & 9.24  & 20.42 & 9.26  & 9.25  & 9.25  & 9.26  & 9.25  & 9.25  & 9.25  & 20.20 & 10.84 & 23.64 & 0     & 4.15  & 7.06  & 5.24  & 2.61  & 17.33 & 6.41  & 28.63 & 0.58  & 5.38  & 8.45  & 8.51  & 8.68  & 9.01  & 1.50  & \cellcolor[rgb]{ .992,  .835,  .498}408.74 \\
    Train & 20211208 & \state{idle}  & 7.99  & 15.18 & 14.57 & 18.94 & 19.50 & 2.00  & 6.78  & 3.82  & 9.53  & 35.73 & 0.07  & 1.48  & 11.02 & 9.25  & 19.89 & 1.38  & 1.32  & 1.33  & 1.38  & 1.33  & 1.33  & 1.33  & 20.07 & 10.85 & 23.84 & 0     & 4.16  & 12.79 & 5.05  & 2.40  & 17.26 & 4.70  & 28.63 & 0.41  & 5.29  & 9.12  & 8.34  & 9.10  & 9.00  & 1.44  & \cellcolor[rgb]{ .98,  .62,  .459}357.63 \\
    \rowcolor[rgb]{ .851,  .851,  .851} Train & 20211209 & \state{idle}  & 7.63  & 14.66 & 14.43 & 18.91 & 19.18 & 1.97  & 6.89  & 3.82  & 9.53  & 35.43 & 0.07  & 1.49  & 11.30 & 9.25  & 20.10 & 9.26  & 9.24  & 9.25  & 9.25  & 9.25  & 9.25  & 9.25  & 19.11 & 10.85 & 23.40 & 0     & 4.16  & 14.62 & 5.21  & 2.54  & 17.30 & 4.70  & 28.63 & 0.65  & 23.08 & 8.29  & 8.58  & 8.39  & 8.70  & 1.42  & \cellcolor[rgb]{ 1,  .922,  .518}429.03 \\
    Train & 20211210 & \state{idle}  & 8.12  & 15.42 & 14.77 & 19.13 & 19.35 & 2.28  & 6.90  & 3.79  & 9.52  & 35.36 & 0.04  & 1.48  & 10.94 & 9.24  & 21.11 & 9.27  & 9.24  & 9.25  & 9.26  & 9.25  & 9.25  & 9.25  & 18.72 & 10.84 & 23.33 & 0     & 4.15  & 23.02 & 5.20  & 2.37  & 17.25 & 4.69  & 28.59 & 0.47  & 6.86  & 8.31  & 8.55  & 8.54  & 8.79  & 1.43  & \cellcolor[rgb]{ .996,  .898,  .51}423.30 \\
    \rowcolor[rgb]{ .851,  .851,  .851} Test  & 20211213 & \state{idle}  & 8.47  & 15.19 & 15.03 & 19.43 & 19.54 & 2.05  & 6.74  & 3.82  & 9.53  & 35.38 & 0.07  & 1.48  & 11.03 & 9.25  & 20.22 & 9.28  & 9.24  & 9.25  & 9.26  & 9.25  & 9.25  & 9.25  & 18.82 & 10.84 & 23.72 & 0     & 4.16  & 20.58 & 5.42  & 2.42  & 17.26 & 4.66  & 28.54 & 0.42  & 6.90  & 8.20  & 8.62  & 8.40  & 8.88  & 1.42  & \cellcolor[rgb]{ .996,  .886,  .51}421.23 \\
    Test  & 20211215 & \state{idle}  & 8.09  & 14.14 & 13.82 & 18.83 & 17.68 & 2.00  & 7.36  & 3.81  & 9.53  & 35.47 & 0.07  & 1.47  & 11.43 & 9.24  & 27.99 & 9.26  & 9.24  & 9.26  & 9.24  & 9.25  & 9.25  & 9.25  & 18.67 & 10.84 & 23.40 & 0     & 4.15  & 21.63 & 5.16  & 2.63  & 17.27 & 4.66  & 28.60 & 0.50  & 5.38  & 8.29  & 8.95  & 8.27  & 9.07  & 1.44  & \cellcolor[rgb]{ .996,  .902,  .514}424.56 \\
    \rowcolor[rgb]{ .851,  .851,  .851} Test  & 20211216 & \state{idle}  & 7.65  & 12.23 & 12.27 & 18.83 & 17.25 & 2.00  & 6.91  & 3.80  & 9.52  & 35.58 & 0.11  & 1.46  & 11.01 & 9.25  & 21.12 & 9.26  & 9.24  & 9.25  & 9.25  & 9.25  & 9.25  & 9.25  & 18.68 & 10.85 & 23.40 & 0     & 4.15  & 25.21 & 5.15  & 2.46  & 17.20 & 4.66  & 28.61 & 0.37  & 5.33  & 8.28  & 9.13  & 8.32  & 9.11  & 1.43  & \cellcolor[rgb]{ .996,  .867,  .506}416.07 \\
    Test  & 20211220 & \state{idle}  & 7.78  & 15.73 & 14.76 & 18.35 & 19.86 & 2.00  & 6.80  & 3.83  & 9.52  & 35.51 & 0.07  & 1.47  & 11.01 & 9.24  & 27.55 & 9.26  & 9.24  & 9.24  & 9.25  & 9.25  & 9.25  & 9.24  & 18.66 & 10.84 & 23.18 & 0     & 4.14  & 21.32 & 5.18  & 2.28  & 17.32 & 4.75  & 28.62 & 0.52  & 5.32  & 9.36  & 9.15  & 9.24  & 8.67  & 1.58  & \cellcolor[rgb]{ .996,  .918,  .514}428.31 \\
    \rowcolor[rgb]{ .851,  .851,  .851} Test  & 20211222 & \state{idle}  & 7.63  & 16.77 & 15.48 & 18.56 & 20.42 & 2.00  & 6.78  & 3.79  & 9.52  & 35.56 & 0.07  & 1.46  & 11.02 & 9.24  & 27.83 & 9.26  & 9.24  & 9.24  & 9.25  & 9.25  & 9.25  & 9.24  & 18.77 & 10.84 & 23.18 & 0     & 4.16  & 22.92 & 5.20  & 2.53  & 17.28 & 4.69  & 28.64 & 2.39  & 6.84  & 9.37  & 9.12  & 9.28  & 8.70  & 1.44  & \cellcolor[rgb]{ 1,  .922,  .518}436.18 \\
    Test  & 20211223 & \state{idle}  & 8.32  & 16.74 & 15.62 & 18.37 & 20.51 & 1.98  & 6.84  & 3.80  & 9.52  & 35.50 & 0.07  & 1.46  & 11.24 & 9.24  & 27.58 & 9.25  & 9.24  & 9.24  & 9.25  & 9.25  & 9.25  & 9.24  & 18.83 & 10.85 & 23.19 & 0     & 4.16  & 20.25 & 5.19  & 2.54  & 17.28 & 4.70  & 28.62 & 3.19  & 6.87  & 9.36  & 9.19  & 9.25  & 8.68  & 1.56  & \cellcolor[rgb]{ 1,  .922,  .518}435.22 \\
    \bottomrule				
    \end{tabular}}%
  \label{tab:packets}%
\end{table}%

\end{landscape}

\begin{table}[htbp]
\setlength{\tabcolsep}{5pt}
  \centering
  \caption{Class-based F1 score results for AA case, showing means of 100 repeats.}
    \begin{tabular}{@{}lrrrr@{}}
    \toprule
    Devices & \multicolumn{1}{l}{precision} & \multicolumn{1}{l}{recall} & \multicolumn{1}{l}{f1-score} & \multicolumn{1}{l}{support} \\
    \midrule
     Amazon AE Dot & \cellcolor[rgb]{ .984,  .98,  .992}0.995 & \cellcolor[rgb]{ .988,  .988,  1}1.000 & \cellcolor[rgb]{ .984,  .984,  .996}0.997 & 22921 \\
    Amazon AE Spot & \cellcolor[rgb]{ .973,  .412,  .42}0.388 & \cellcolor[rgb]{ .984,  .984,  .996}0.999 & \cellcolor[rgb]{ .98,  .702,  .714}0.559 & 15265 \\
    Amazon AE Studio & \cellcolor[rgb]{ .984,  .953,  .965}0.963 & \cellcolor[rgb]{ .988,  .988,  1}1.000 & \cellcolor[rgb]{ .984,  .973,  .984}0.981 & 14515 \\
    Amazon Plug & \cellcolor[rgb]{ .988,  .988,  1}1.000 & \cellcolor[rgb]{ .984,  .984,  .996}0.999 & \cellcolor[rgb]{ .988,  .988,  1}1.000 & 1392 \\
    Amcrest WiFi-Cam. & \cellcolor[rgb]{ .984,  .973,  .984}0.984 & \cellcolor[rgb]{ .988,  .988,  1}1.000 & \cellcolor[rgb]{ .984,  .98,  .992}0.992 & 5978 \\
    Arlo Base Station & \cellcolor[rgb]{ .988,  .988,  1}1.000 & \cellcolor[rgb]{ .988,  .988,  1}1.000 & \cellcolor[rgb]{ .988,  .988,  1}1.000 & 84231 \\
    Arlo Q Camera & \cellcolor[rgb]{ .988,  .988,  1}1.000 & \cellcolor[rgb]{ .988,  .988,  1}1.000 & \cellcolor[rgb]{ .988,  .988,  1}1.000 & 43252 \\
    Atomi Coff-Maker & \cellcolor[rgb]{ .988,  .988,  1}1.000 & \cellcolor[rgb]{ .984,  .984,  .996}0.997 & \cellcolor[rgb]{ .984,  .984,  .996}0.999 & 7263 \\
    Borun Camera & \cellcolor[rgb]{ .984,  .984,  .996}0.998 & \cellcolor[rgb]{ .988,  .988,  1}1.000 & \cellcolor[rgb]{ .984,  .984,  .996}0.999 & 26338 \\
    D-Link Mini Cam. & \cellcolor[rgb]{ .988,  .988,  1}1.000 & \cellcolor[rgb]{ .988,  .988,  1}1.000 & \cellcolor[rgb]{ .988,  .988,  1}1.000 & 1158 \\
    D-Link Water Sen. & \cellcolor[rgb]{ .988,  .988,  1}1.000 & \cellcolor[rgb]{ .988,  .988,  1}1.000 & \cellcolor[rgb]{ .988,  .988,  1}1.000 & 50 \\
    Eufy HomeBase 2 & \cellcolor[rgb]{ .988,  .988,  1}1.000 & \cellcolor[rgb]{ .988,  .988,  1}1.000 & \cellcolor[rgb]{ .988,  .988,  1}1.000 & 8304 \\
    Globe Lamp & \cellcolor[rgb]{ .984,  .98,  .992}0.995 & \cellcolor[rgb]{ .984,  .878,  .886}0.821 & \cellcolor[rgb]{ .984,  .922,  .933}0.900 & 6985 \\
    Google Nest Mini & \cellcolor[rgb]{ .988,  .988,  1}1.000 & \cellcolor[rgb]{ .988,  .988,  1}1.000 & \cellcolor[rgb]{ .988,  .988,  1}1.000 & 19711 \\
    Gosund Plug & \cellcolor[rgb]{ .984,  .929,  .941}0.939 & \cellcolor[rgb]{ .984,  .984,  .996}0.999 & \cellcolor[rgb]{ .984,  .965,  .976}0.968 & 24042 \\
    Gosund Socket & \cellcolor[rgb]{ .988,  .988,  1}1.000 & \cellcolor[rgb]{ .984,  .976,  .988}0.984 & \cellcolor[rgb]{ .984,  .98,  .992}0.992 & 18355 \\
    HeimVision Lamp & \cellcolor[rgb]{ .988,  .988,  1}1.000 & \cellcolor[rgb]{ .984,  .945,  .957}0.932 & \cellcolor[rgb]{ .984,  .965,  .976}0.965 & 8303 \\
    HeimVision S Cam. & \cellcolor[rgb]{ .988,  .988,  1}1.000 & \cellcolor[rgb]{ .988,  .988,  1}1.000 & \cellcolor[rgb]{ .988,  .988,  1}1.000 & 12539 \\
    Home Eye Camera & \cellcolor[rgb]{ .988,  .988,  1}1.000 & \cellcolor[rgb]{ .988,  .988,  1}1.000 & \cellcolor[rgb]{ .988,  .988,  1}1.000 & 26253 \\
    iRobot Roomba & \cellcolor[rgb]{ .988,  .988,  1}1.000 & \cellcolor[rgb]{ .988,  .988,  1}1.000 & \cellcolor[rgb]{ .988,  .988,  1}1.000 & 1136 \\
    Luohe Cam Dog & \cellcolor[rgb]{ .988,  .988,  1}1.000 & \cellcolor[rgb]{ .988,  .988,  1}1.000 & \cellcolor[rgb]{ .988,  .988,  1}1.000 & 3171 \\
    Nest Indoor Cam. & \cellcolor[rgb]{ .988,  .988,  1}1.000 & \cellcolor[rgb]{ .984,  .984,  .996}0.999 & \cellcolor[rgb]{ .984,  .984,  .996}0.999 & 48001 \\
    Netatmo Camera & \cellcolor[rgb]{ .988,  .988,  1}1.000 & \cellcolor[rgb]{ .988,  .988,  1}1.000 & \cellcolor[rgb]{ .988,  .988,  1}1.000 & 9059 \\
    Netatmo Weather & \cellcolor[rgb]{ .988,  .988,  1}1.000 & \cellcolor[rgb]{ .988,  .988,  1}1.000 & \cellcolor[rgb]{ .988,  .988,  1}1.000 & 1912 \\
    Philips Hue Bridge & \cellcolor[rgb]{ .988,  .988,  1}1.000 & \cellcolor[rgb]{ .988,  .988,  1}1.000 & \cellcolor[rgb]{ .988,  .988,  1}1.000 & 13608 \\
    Ring Base Station & \cellcolor[rgb]{ .988,  .988,  1}1.000 & \cellcolor[rgb]{ .973,  .51,  .522}0.223 & \cellcolor[rgb]{ .976,  .557,  .569}0.336 & 3732 \\
    SIMCAM 1S & \cellcolor[rgb]{ .988,  .988,  1}1.000 & \cellcolor[rgb]{ .988,  .988,  1}1.000 & \cellcolor[rgb]{ .988,  .988,  1}1.000 & 23660 \\
    Smart Board & \cellcolor[rgb]{ .98,  .737,  .745}0.735 & \cellcolor[rgb]{ .973,  .412,  .42}0.057 & \cellcolor[rgb]{ .973,  .412,  .42}0.105 & 24021 \\
    Sonos One Speaker & \cellcolor[rgb]{ .984,  .969,  .98}0.983 & \cellcolor[rgb]{ .988,  .988,  1}1.000 & \cellcolor[rgb]{ .984,  .98,  .992}0.991 & 5074 \\
    Teckin Plug & \cellcolor[rgb]{ .984,  .945,  .957}0.956 & \cellcolor[rgb]{ .984,  .929,  .941}0.909 & \cellcolor[rgb]{ .984,  .941,  .953}0.932 & 12847 \\
    Yutron Plug & \cellcolor[rgb]{ .984,  .906,  .918}0.916 & \cellcolor[rgb]{ .988,  .988,  1}1.000 & \cellcolor[rgb]{ .984,  .957,  .969}0.956 & 13233 \\

    \midrule
    accuracy & \multicolumn{4}{c}{0.943} \\
    macro avg & 0.963 & 0.933 & 0.925 & 506309 \\
    weighted avg & 0.961 & 0.943 & 0.932 & 506309 \\
    \hline
    \end{tabular}%
  \label{tab:class-aa}%
\end{table}%

\end{document}